\documentclass[useAMS,fleqn,usenatbib]{mnras}

\usepackage[T1]{fontenc}
\usepackage{ae,aecompl}

\usepackage{amsmath}
\usepackage{amsopn}
\usepackage[british]{babel}
\usepackage[varg]{txfonts}
\usepackage{biblio} 
\usepackage{natbib}
\usepackage{color}
\usepackage{bm}
\usepackage[normalem]{ulem}
\usepackage{graphicx}
\usepackage{epstopdf } 

\usepackage{microtype}

\usepackage{bold-extra}
\definecolor{orange}{rgb}{1.0,0.5,0.}

\def\fdisc{\ifmmode{\>f_{\rm disc}}\else{$f_{\rm disc}$}\fi}
\def\Rd{\ifmmode{\>R_{\rm d}}\else{$R_{\rm d}$}\fi}
\def\md{\ifmmode{\>m_{\rm d}}\else{$m_{\rm d}$}\fi}

\def\MDM{\ifmmode{\>M_{\textnormal{\sc dm}}}\else{$M_{\textnormal{\sc dm}}$}\fi}

\def\XH{\ifmmode{\>X_{\textnormal{\sc h}}} \else{$X_{\textnormal{\sc h}}$}\fi}
\def\nH{\ifmmode{\>n_{\textnormal{\sc h}}} \else{$n_{\textnormal{\sc h}}$}\fi}

\def\maspyr{\ifmmode{\>\textnormal{mas~yr}^{-1}}\else{mas~yr$^{-1}$}\fi}

\def\mG{\ifmmode{\>\mu\mathrm{G}}\else{$\mu$G}\fi}
\def\erg{\ifmmode{\> {\rm erg}}\else{erg}\fi}
\def\keV{\ifmmode{\> {\rm keV}}\else{keV}\fi}

\def\deg{\ifmmode{\>^{\circ}}\else{$^{\circ}$}\fi}
\def\onedeg{\ifmmode{\>1^{\circ}}\else{$1^{\circ}$}\fi}

\def\xvir{\ifmmode{\>\!x_{vir}}\else{$x_{vir}$}\fi}
\def\Mvir{\ifmmode{\>\!M_{vir} }\else{$M_{vir} $}\fi}
\def\rvir{\ifmmode{\>\!r_{vir}}\else{$r_{vir}$}\fi}
\def\vvir{\ifmmode{\>\!v_{vir}}\else{$v_{vir}$}\fi}
\def\Vvir{\ifmmode{\>\!V_{vir} }\else{$V_{vir} $}\fi}

\def\tratio{\ifmmode{\>\tau}\else{$\tau$}\fi}

\def\rms{\ifmmode{\>r_{\textnormal{\sc ms}}}\else{$r_{\textnormal{\sc ms}}$}\fi}

\def\Mpc{\ifmmode{\>\!{\rm Mpc}} \else{Mpc}\fi}
\def\kpc{\ifmmode{\>\!{\rm kpc}} \else{kpc}\fi}
\def\pc{\ifmmode{\>\!{\rm pc}} \else{pc}\fi}

\def\Gyr{\ifmmode{\>\!{\rm Gyr}} \else{Gyr}\fi}
\def\Myr{\ifmmode{\>\!{\rm Myr}} \else{Myr}\fi}
\def\yr{\ifmmode{\>\!{\rm yr}} \else{yr}\fi}
\def\pyr{\ifmmode{\>\!{\rm yr}^{-1}}\else{yr $^{-1}$} \fi}
\def\s{\ifmmode{\>\!{\rm s}}\else{s}\fi}
\def\ps{\ifmmode{\>\!{\rm s}^{-1}}\else{s$^{-1}$}\fi}
\def\Hz{\ifmmode{\>\!{\rm Hz}}\else{Hz}\fi}

\def\kms{\ifmmode{\>\!{\rm km\,s}^{-1}}\else{km~s$^{-1}$}\fi}

\def\K{\ifmmode{\>\!{\rm K}}\else{K}\fi}

\def\sr{\ifmmode{\>\!{\rm sr}}\else{sr}\fi}
\def\psr{\ifmmode{\>\!{\rm sr}^{-1}}\else{sr$^{-1}$}\fi}
\def\arcs{\ifmmode{\>\!{\rm arcsec}}\else{arcsec}\fi}
\def\parcs{\ifmmode{\>\!{\rm arcsec}^{-1}}\else{arcsec${-1}$}\fi}
\def\parcss{\ifmmode{\>\!{\rm arcsec}^{-2}}\else{arcsec${-2}$}\fi}

\def\cm{\ifmmode{\>\!{\rm cm}}\else{cm}\fi}
\def\cc{\ifmmode{\>\!{\rm cm}^{3}}\else{cm$^{3}$}\fi}
\def\sqc{\ifmmode{\>\!{\rm cm}^{2}}\else{cm$^{2}$}\fi}
\def\pcc{\ifmmode{\>\!{\rm cm}^{-3}}\else{cm$^{-3}$}\fi}
\def\psc{\ifmmode{\>\!{\rm cm}^{-2}}\else{cm$^{-2}$}\fi}

\def\g{\ifmmode{\>\!{\rm g}}\else{g}\fi}
\def\Msun{\ifmmode{\>\!{\rm M}_{\odot}}\else{M$_{\odot}$}\fi}
\def\hMsun{\ifmmode{\> h^{-1}{\rm M}_{\odot}}\else{$h^{-1}$M$_{\odot}$}\fi}

\def\Zsun{\ifmmode{\>\!{\rm Z}_{\odot}}\else{Z$_{\odot}$}\fi}

\def\Lsun{\ifmmode{\>\!{\rm L}_{\odot}}\else{L$_{\odot}$}\fi}

\def\rayl{\ifmmode{\>\!{\rm R}}\else{R}\fi}
\def\mR{\ifmmode{\>\!{\rm mR}}\else{mR}\fi}

\renewcommand{\ion}[2]{\hbox{#1\,{\sc #2}}}

\def\lya{\ifmmode{\>\!{\rm Ly}\alpha}\else{Ly$\alpha$}\fi}

\def\Ha{\ifmmode{\>\!{\rm H}\alpha}\else{H$\alpha$}\fi}
\def\Hb{\ifmmode{\>\!{\rm H}\beta}\else{H$\beta$}\fi}

\def\HI{\ifmmode{\> \textnormal{\ion{H}{i}}} \else{\ion{H}{i}}\fi}
\def\HII{\ifmmode{\> \textnormal{\ion{H}{ii}}} \else{\ion{H}{ii}}\fi}
\def\CIV{\ifmmode{\> \textnormal{\ion{C}{iv}}} \else{\ion{C}{iv}}\fi}
\def\SiIV{\ifmmode{\> \textnormal{\ion{S}{iv}}} \else{\ion{Si}{iv}}\fi}

\def\NH{\ifmmode{\> {\rm N}_{\rm H}} \else{N$_{\rm H}$}\fi}
\def\Ng{\ifmmode{\> {\rm N}_{\rm gas}} \else{N$_{\rm gas}$}\fi}
\def\NHI{\ifmmode{\> {\rm N}_{\HI}} \else{N$_{\HI}$}\fi}
\def\MHI{\ifmmode{\> {\rm M}_{ \HI}} \else{M$_{\HI}$}\fi}

\def\mua{\ifmmode{\>\mu_{ \textnormal{\Ha}}}\else{$\mu_{ \textnormal{\Ha}}$}\fi}
\def\alphabha{\ifmmode{\>\alpha_{B}^{(\textnormal{\Ha})}}\else{$\alpha_{B}^{(\textnormal{\Ha})}$}\fi}

\newcommand{\myemail}{tepper@physics.usyd.edu.au}
\newcommand{\ramses}{{\sc Ramses}}
\newcommand{\agama}{{\small AGAMA}}
\newcommand{\pynbody}{{\sc Pynbody}}

\newcommand{\nexus}{{\sc Nexus}}
\newcommand{\dice}{{\small DICE}}

\defcitealias{mo98a}{MMW98}

\title[ \nexus-{\sc cdm}: Method and Validation ]{ {\bfseries \textsc{Nexus - cdm}}: Isolated Galaxy Simulations with Cosmologically Evolving Dark-Matter Halos  I. Method and Validation }

\author[Tepper-Garc\'ia et al.]{%
Thor Tepper-Garc\'ia$,^{1}$\thanks{\myemail} 
Joss Bland-Hawthorn,$^{1}$
Oscar Agertz,$^{2}$ and
Christoph Federrath$^{3}$
\\
$^{1}$Sydney Institute for Astronomy, School of Physics, A28, The University of Sydney, NSW 2006, Australia\\
$^{2}$Lund Observatory, Division of Astrophysics, Department of Physics, Lund University, Box 118, SE-221 00 Lund, Sweden\\
$^{3}$Research School of Astronomy and Astrophysics, Australian National University, Canberra, ACT 2611, Australia
}

\date{Accepted ---. Received ---; in original form ---}

\pubyear{\date{year}}

\begin{document}
\label{firstpage}
\pagerange{\pageref{firstpage}--\pageref{lastpage}}
\maketitle

\begin{abstract}
Galaxy evolution can be modelled in two complementary ways.
Standalone simulations place a single galaxy in isolation, evolving it under controlled initial conditions without the surrounding cosmic web. 
This approach offers high resolution and computational efficiency, making it well suited to disentangle specific physical processes (e.g.~feedback and disc instabilities) and testing them cleanly, though at the cost of ignoring environmental context, gas accretion from the broader large-scale structure, and the hierarchical assembly that feeds real galaxies over cosmic time. Cosmological simulations, by contrast, form galaxies self-consistently within an expanding universe, capturing large-scale structure formation, mergers,
gas inflows, and environmental effects such as tidal stripping and ram-pressure. This realism comes at a steep computational cost, requiring coarser resolution or simplified sub-grid physics for star formation and feedback. Here, we explore a new paradigm that combines the relative merits of both methods -- what we call the {\sc Nexus-\small CDM} framework. This approach naturally incorporates galaxy mass growth, which is absent from existing standalone simulations,
reinforcing the limitations of static halo models used in galaxy formation and evolution research.
The viability of the framework is demonstrated in an idealised setup, where it successfully reproduces the formation and subsequent evolution of a realistic stellar disc.
We present some early results, in particular, discs form readily in a gravitational potential with a shallow central gradient, contrary to recent claims.

\end{abstract}


\section{Introduction} \label{sec:intro}

The current paradigm of galaxy formation postulates that galaxies form within the potential well of dark matter (DM) halos, which themselves form by hierarchical merging of smaller halos \citep{whi78a,pre74a}, leading to the rich hierarchy of structures -- from galaxies and groups to clusters, filaments, walls, and voids -- observed across cosmic time \citep{coll01,abd25a}.
The processes of hierarchical large-scale structure formation, and the emergence of galaxies embedded within, has been well demonstrated by complex N-body gas-dynamical computer simulations \citep[a.k.a. `cosmological' simulations; e.g.][]{sch15a,pil18l,dub21a,age21l}.\footnote{ See \citealt{naa17a} for a review of the state-of-the-art up to a decade ago. }

One drawback of cosmological simulations is that, by design, they offer limited control over the structural and kinematic state of a galaxy at a given epoch; indeed, the whole point of such simulations is to emulate the formation and evolution of galaxies from first principles, in particular, from the density fluctuations present in the infant Universe \citep{zel70a}. 

A different approach is that offered by controlled simulations of isolated (or standalone) galaxies, which can be traced back to early simulations by 
\cite{lind61} and \citet{hoh71a}. In plain terms, this type of simulation entails the study of the formation and/or evolution of `a galaxy in a box',
i.e. a system that evolves from well-defined, but {\it ad hoc}, initial conditions, and does so in the absence of any cosmological boundary conditions, notably the inflow of matter (e.g.~gas streams and DM sub-halos), but also of the gravitational effect of the environment (groups, clusters, filaments) in which galaxies typically form. The galaxy may be subject to some type of interaction, as first demonstrated in early simulations by \citet{hol41a} and \citet{too72a},
but all of this is carried out in a controlled manner within an isolated environment. Standalone galaxy simulations have a remarkable history of identifying important dynamical processes $-$ bar formation, tidal features, disc warps, disc/bar buckling, stellar migration, etc. \citep[q.v.][see their sec.~7]{tep24a}.

Sitting between these two contrasting approaches is a technique based on `genetic modification' (GM) of the initial conditions, pioneered by \citet{rot16a} and refined by \citet{rey18a}. GM simulations, which build upon the `zoom-in' approach \citep{kat93b,kat94a}, allow for some control over the conditions under which a galaxy evolves by systematically varying the mass and assembly timeline of its host DM halo -- extracted from a full cosmological simulation -- with the aim to explore how each of these variations affects the galaxy's properties \citep[e.g.][]{rey2023,joshi2024}. A drawback of this technique is that, while it is possible to alter the linear perturbation field to predictably achieve a later (or an earlier) growth of a halo or a different final mass, non-linear effects also play a role, making it difficult to control the details of the merger history.

Another approach offering some control over a simulation relies on the use of an analytic reconstruction of a cosmologically evolving DM halo and its gravitational potential. This idea was first investigated by \citet{low11a} but its basis dates back to self-consistent field methods described by \citet{clu73} and \citet{her92a}. The method has
since been employed by several follow-up studies \citep[e.g.][]{nga15a,san20b,aro22a}. The essence of the method is to approximate the potential of the N-body DM halo with an analytic function; the halo's evolution is captured by carrying out the potential reconstruction at a number of selected points in time over the relevant time span of the simulation. The general intent is to replace the full, N-body DM halo by its analytic representation, and to `replay'  the simulation or, more commonly, some variation thereof, at the same (or at a significantly higher) resolution than the original simulation. 

The most widespread application of this technique is the dynamical modelling and systematic study of a specific sub-system evolving within the DM halo, such as: the orbits of sub-halos \citep{low11a}; satellite configurations \citep{san20b}; the formation and structure of tidal streams \citep{nga15a} and stellar tidal debris \citep{aro25a}. Other applications consist of the study of the host halo's shape distorted by its environment, in particular by a massive satellite, both within cosmological boundary conditions \citep{ara25a} and isolated galaxy models \citep{gar21a}.

In this approach, the analytic reconstruction of the potential relies on the use of some form of basis-function series expansion, where the radial dependence is split from the angular dependence, and where both dependencies are generally described by independent orthonormal basis sets. Given the generic, quasi-spheroidal structure (although intrinsically triaxial) of DM halos as found in simulations over a wide range of environmental conditions and epochs \citep[][]{dub91a,all06a}, the use of spherical harmonics is the natural choice to describe the potential's angular dependence. The choice of radial basis functions generally comes down to two options. On the one hand, a radial basis function set is derived from the \citet{her90a} profile as developed by \citet[][]{her92a}. Such a set has been adopted by \citet{low11a} and \citet{san20b}. On the other hand, the basis radial set consists of interpolating functions (splines) of a prescribed order defined on a radial grid \citep{vas13h}, an approach adopted by \citet{san20b}, \citet{aro22a} and \citet{ara25a}.\\

So far, it seems, the application of the `surrogate halo'  simulation technique -- as we refer to it -- described above has focused on {\em gas-free} systems. A natural extension of this practice is to consider the presence of a gaseous component when replaying a simulation. Our specific goal here is to explore such an extension.\footnote{See \citet{aum13a} for a related approach, but using fully responsive DM halos.}

In a complementary manner to the GM technique -- where the host DM halo properties are systematically altered -- our intention relies on fixing the evolution of the DM halo, while systematically varying the initial conditions of the baryons evolving within.
In addition, we wish to explore how, for specific baryonic initial conditions, different numerical ingredients (sub-grid recipes, resolution, etc.), and the values of their physically unconstrained parameters, affect the galaxy's evolution, potentially at a lesser computational cost.
The end goal is to extend our \nexus\ framework\footnote{\url{https://www.physics.usyd.edu.au/nexus/}} \citep{tep24a} to enable controlled simulations of individual galaxies embedded within a cosmologically evolving host DM halo,
as {\em a further step towards bridging the gap between theory, full cosmological simulations and isolated galaxy simulations}. We dub this extended framework \nexus-{\sc cdm}.

Owing to the substantial amount of work involved in this endeavour, we present our results across several papers. In this first paper of the series, we focus on introducing our method and its application to a first, simple, but crucial validation test.
We broach this with the following steps:
\begin{itemize}
    \item[] Step 1. Run an idealised simulation of a standalone galaxy forming and evolving within a non-cosmological, {\em fully responsive} DM halo (benchmark simulation).
    \item[] Step 2. Reconstruct the DM halo's evolution with an analytic model (surrogate halo).
    \item[] Step 3. Replay the simulation with the surrogate halo in place of the fully responsive DM halo (surrogate halo simulation).
    \item[] Step 4. Compare the evolution and final properties of the galaxy in the benchmark and surrogate halo simulations.
\end{itemize}
The objective here is to explore whether the surrogate halo simulation recovers the evolution and generic properties of the galaxy that forms in the benchmark simulation with sufficient fidelity. A positive outcome would open the door to applying this technique to halos extracted from cosmological simulations -- which is our ultimate goal -- and thus justify subsequent papers in this series.\\

This first paper is structured according to the steps given above as follows. We start by introducing our method, providing details about the benchmark simulation employed to calculate the evolution of the DM halo and the galaxy within. Then, we introduce our approach to reconstruct the DM halo's potential, and follow with an evaluation of this reconstruction. Next, we introduce our surrogate halo simulation, as well as a number of additional simulations in which the fully responsive DM halo potential is replaced by different analytic reconstructions. We follow with a comparative analysis of the structural and kinematic properties of the disc that forms in each of these simulations. We conclude by summarising our most important findings, by discussing caveats of our approach, and by providing an outlook of the work to come.

\section{Method} \label{sec:meth}

In a nutshell, our intent is to model the formation and evolution of a disc galaxy within a fully responsive, isolated DM halo, and to explore whether the halo's potential can be reconstructed analytically with sufficient fidelity to replay the simulation and recover the evolution of the stellar disc, and of its gaseous component, that form within.

The first step is thus to setup and run a fully self-consistent simulation of the formation of a disc galaxy, but with the simplest configuration possible, that will serve as a benchmark against which to test the analytic potential reconstruction. In other words, this experiment is not intended to span the full diversity of galaxy formation pathways, but to isolate whether the evolving gravitational field of a cosmological halo can be accurately represented by a surrogate halo.

\subsection{Benchmark Simulation} \label{sec:bench}

\begin{table}
\begin{center}
\caption{Model parameters in the benchmark run (Sec.~\ref{sec:bench}). The total mass, scale length and virial radius are indicated in columns 2, 3, and 4, respectively. Column~5 gives the number of particles used to sample the corresponding component.
}
\label{tab:bench}
\begin{tabular}{llccccc}
Component & Mass & $r_{\rm s}$ & $r_{\rm vir}$ & N \\
 & ($10^{10}$ \Msun) & (kpc) & (kpc) & ($10^6$) \\
\hline
DM halo & 50 & 20.8 & 121.7& 1 \\
Hot halo & 2.5 & 20.8 & 121.7 & 2 \\
\hline
\end{tabular}
\end{center}
\begin{list}{}{}
\item Notes. Both components follow a \citet[][NFW]{nav97a} density profile. The spin parameter of both halos is set to $\lambda = 0.12$. The initial metallicity is set to $5\times10^{-2}~\Zsun$. The density profiles, the derived rotation velocity profiles, and the gas temperature profile are shown in Fig.~\ref{fig:halos}.
\end{list}
\end{table}

\begin{figure*}
\centering
\includegraphics[width=0.33\textwidth]{ 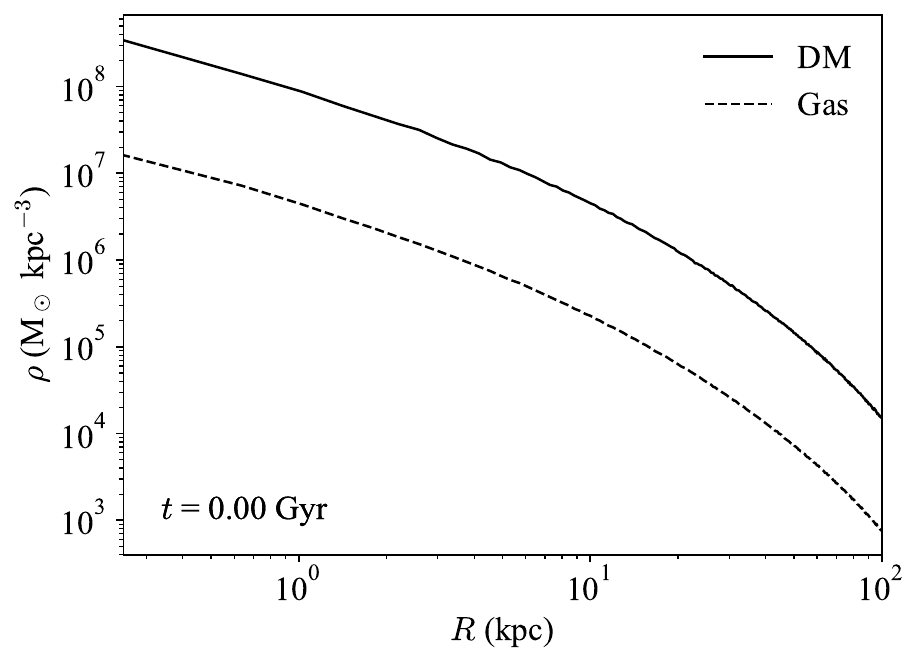 }
\includegraphics[width=0.33\textwidth]{ 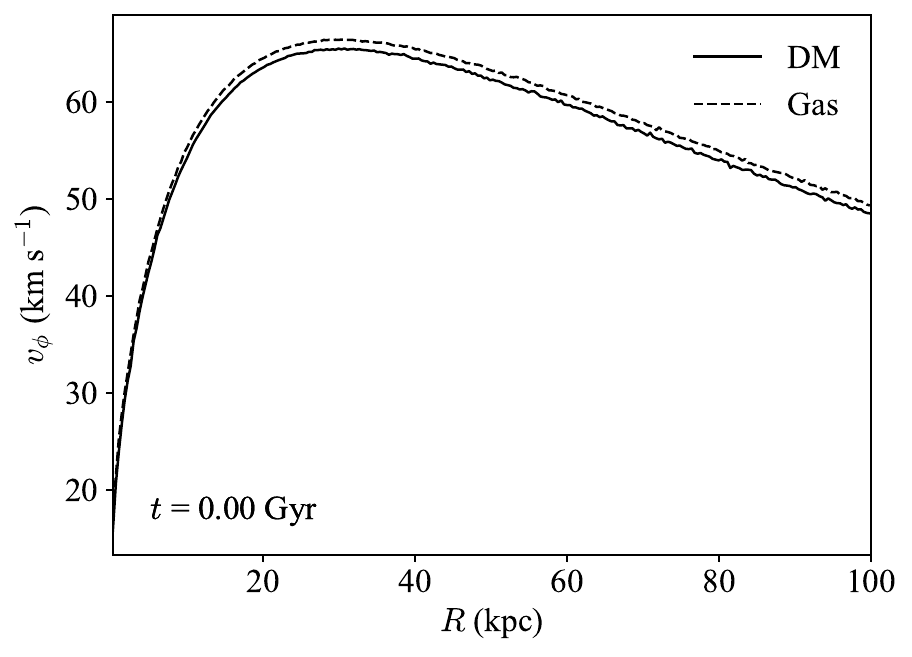 }
\includegraphics[width=0.33\textwidth]{ 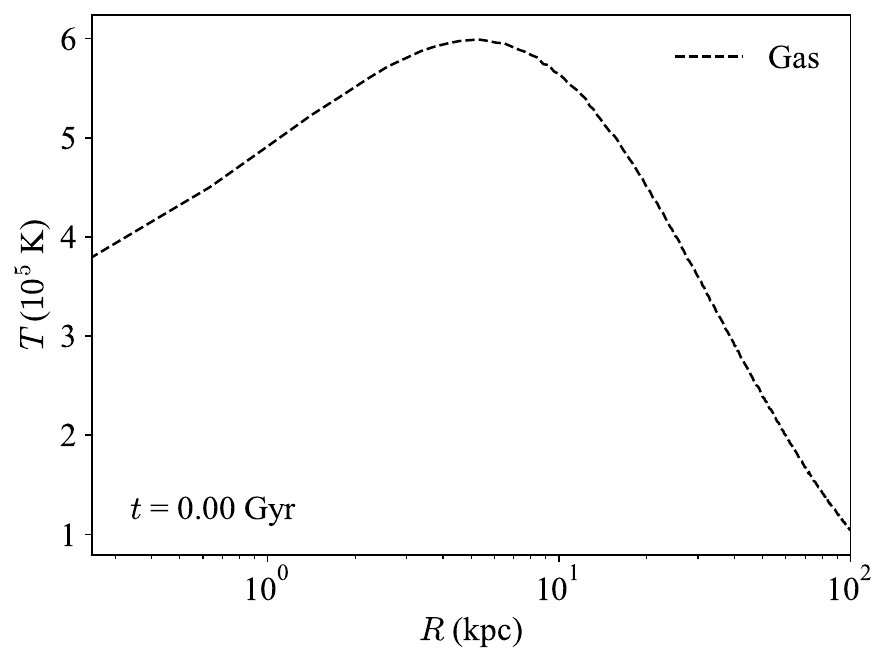 }
\caption{Initial properties of the DM halo (solid curves) and hot gas halo (dashed curves). Left: Volume density profile. Centre: Rotational velocity profile. Right: Gas temperature profile. All quantities correspond to azimuthal averages as a function of cylindrical radius $R$. Note that the specific angular momenta of the DM and gas components are equal by design, but that of the gas has been displaced vertically by 1~\kms\ for visualisation purposes. The DM halo has a virial mass \mbox{$M_{\textnormal{\sc dm}} \approx 5\times10^{11}$~\Msun} within $r_{\rm vir} \approx 122$~kpc, and the hot gas halo a total mass \mbox{$M_{\rm g} \approx 2.5\times10^{10}$~\Msun} within the same radius. The gas-to-total mass fraction is thus roughly 5~percent.
}
\label{fig:halos}
\end{figure*}

Our benchmark run is similar to the `cooling halo' setup presented as part of the validation tests of the \nexus\ framework \citep[][their sec.~5.1.3]{tep24a}. In brief, the system consists initially of a hot (\mbox{$T \sim 5\times10^5$~K}), spinning (\mbox{$v_{\rm rot} \sim 20 - 70$~\kms}) gas halo with mass \mbox{$M_{\rm g} \approx 2.5\times10^{10}$~\Msun}, embedded within a spinning (\mbox{$v_{\rm rot} \sim 20 - 70$~\kms}) DM halo with total mass \mbox{$M_{\textnormal{\sc dm}} \approx 5\times10^{11}$~\Msun}. Both components follow initially a \citet[][NFW]{nav97a} density profile with a scale radius $r_{\rm s} \approx 21$~kpc and a virial radius $r_{\rm vir} \approx 122$~kpc. The gas-to-total mass fraction is roughly 5~percent. The hot gas is initially in vertical hydrostatic equilibrium and the initial gas metallicity is set to $5\times10^{-2}~\Zsun$. The DM halo is sampled with $N_{\textnormal{\sc dm}} = 10^6$ particles, and the gas halo, with $N_{\rm gas} = 2\times10^6$ particles. At initialisation, the gas particles are mapped onto an adaptive mesh refinement (AMR) grid, and thereby converted into cells, with a number dictated by the refinement strategy (see details below). Some basic properties of these components are displayed in Fig.~\ref{fig:halos}, and the model parameters summarised in Tab.~\ref{tab:bench}.

The DM halo profile, as well as the structural and kinematical properties of the hot gas halo are guided by the formalism put forward by \citet[][]{mo98a}. In brief, their approach assumes that, upon virialisation of the host halo, dark matter and baryons (in the form of hot gas) are well mixed, that they follow roughly the same mass distribution, and that their specific angular momenta, $j_{\textnormal{\sc dm}}$ and $j_{\rm gas}$, are roughly equal, all prior to the progressive collapse of the hot halo as a result of cooling and the consequent formation of a stellar disc within. The initial profile of the specific angular momentum is guided by cosmological, disipationless simulations of structure formation \citep{bul01b}, which suggest that $j \propto M^s$ with $s = 1.3 \pm 0.3$, where $M$ is the total mass enclosed within $r$. We adopt $s = 1$.

The DM halo mass is motivated by the existence of an apparent halo mass threshold $M_{\rm h} \gtrsim 10^{11}$~\Msun\ for disc formation \citep[][]{pil18l,el-18j,dub21a,sem26a}. The adopted mass is consistent with the total galaxy mass of a MW-like system at a redshift $z_{\rm cosmo} \approx 1$ based on models of cosmological halo growth \citep[cf.][their fig.~1]{bla16a}. The halo concentration and virial radius are self-consistently calculated for the given mass corresponding redshift with help of the {\sc commah} package \citep{cor15h}. Note that the initial metallicity we adopt is appropriate for gas at this redshift \citep[][]{wie09b}. The number of DM particles is chosen based on the well-established result that $N_{\textnormal{\sc dm}} \gtrsim 10^6$ is required to capture the halo's evolution with some degree of fidelity, as well as that of the stellar disc that forms within \citep[][]{lud21p}. For computational convenience, we set the number of particles at the lower end of this estimate ($N_{\textnormal{\sc dm}} = 10^6$).

The presence of a fully formed, spinning, virialised hot gas halo at  $z_{\rm cosmo} \approx 1$  is expected from theoretical arguments \citep{bir03a,dek06a}, and is supported by the results from cosmological simulations (e.g. \citealt{opp18a,sul26}; \textcolor{blue}{Rufo-Pastor et al., in revision}); observations also provide compelling evidence for their existence, albeit so far only at low redshift \citep[for a recent discussion, see][see also \citealt{tep24a}, their sec.~2.1]{bre18a}. Its mass is chosen such that the resulting baryon-to-total mass fraction of the system ensures: 1) disc stability against bar formation \citep{mo98a}, and 2) consistency with the observed baryon budget of galaxies \citep{mcg26a}. The former constraint is relevant to the validity of the method (see Sec.~\ref{sec:cav}).

The initial conditions and their evolution were calculated with \nexus. In brief, particle positions and velocities for the composite system are self-consistently generated with the Action-based Galaxy Modelling Architecture stellar dynamics library \citep[\agama;][]{vas19a}, modified to treat both collisionless- and gas components. The particle velocities (both DM and gas) are post-processed with the \pynbody\ package \citep{pon13a} to satisfy the conditions imposed on the spin of the system as well as the thermodynamic- and chemical properties of the hot halo, as outlined above.

The simulation was run with the AMR code \ramses\ \citep[][]{tey02a}, and the aforementioned initial conditions were loaded into \ramses\ with a modified version of the \dice\ patch \citep[][]{per14c,per16a}. The simulation was evolved for a total timespan of roughly \mbox{2~Gyr} in a cubic box of size \mbox{300~kpc} per side, adopting a maximum (minimum) AMR level \mbox{$l_{\rm max} = 13$} (\mbox{$l_{\rm min} = 7$}), implying a limiting spatial resolution of \mbox{$\delta x = 300 / 2^{13} \approx 37$~pc}. The AMR grid is refined (de-refined) at runtime mainly according to two criteria: 1) the gas mass within the cell exceeds (or drops below) \mbox{$m_{\rm sph} = 5\times10^{-2}$~M$_{\rm cu}$}, where the mass code unit is \mbox{M$_{\rm cu} = 2.32\times10^5$~\Msun}; or 2) the number of collisionless particles, i.e. DM and stars, exceeds (or drops below) $N_{\rm min} = 8$. These two criteria become particularly relevant when the N-body representation of the DM halo is replaced by an analytic model (cf. Sec.~\ref{sec:concl}). We refer the reader to \citet{tep24a} for specific details about the generation of the initial conditions and their evolution, the adopted sub-grid recipes, and further numerical details adopted in \nexus, which are of lesser relevance to this study.

It should be noted that the adopted simulation timespan, albeit arbitrary, is cosmologically relevant, as it is believed that the disc of the Milky Way formed within a timespan of $\sim 1 - 2$~Gyr \citep{bel22a,con22a}. Also, it is long enough to allow for the disc to develop and settle dynamically in our benchmark run (see App.~\ref{app:disc_form}).

During the system's evolution, the gas is allowed to evolve thermodynamically (cool / heat) and to form stars, according to the prescriptions outlined in \citet[][]{age13a}, \citet[][]{age15a} and \citet[][]{age21l}. Stars evolve as well, feeding energy and mass back into their surroundings, and enriching it with heavy elements.
The system evolves away from its initial configuration; the hot gas loses part of its initial pressure support as a result of cooling, and collapses. Angular momentum conservation deters the gas from collapsing spherically, and so the collapse proceeds mainly along the spin axis, resulting in a disc-like, rotationally supported gas configuration, out of which a stellar disc gradually forms.

This cooling halo setup constitutes our benchmark simulation. It is important to emphasise that in this run the DM halo is fully responsive, as are the gas and the stars that form during the system's evolution. In what follows,  we refer to this run interchangeably as either `N-body, responsive halo' run or `benchmark' run.

\subsubsection{Potential Reconstruction} \label{sec:rec}

Our aim is not to model the formation and evolution of a galactic disc down to every detail, but rather to explore to which extent we can recover its evolution {\it as given by the benchmark run}, when the potential induced by the responsive DM component is replaced by an analytic reconstruction.

To arrive at the latter, we exploit the machinery of \agama, which allows to construct an estimate of the potential, $\Phi_{\rm ana}$, directly from an N-body snapshot (taken at time $t$) as a sum of spherical-harmonic functions of angles multiplied by arbitrary functions of radius,
\begin{equation} \label{eq:phi_ana}
	\Phi_{\rm ana}(r, \psi, \phi, t) = \sum_{\ell,m} \tilde{\Phi}_{\ell,m}(r, t)~Y_\ell^m(\psi, \phi, t) \, ,
\end{equation}
The radial dependence of each term is given by a quintic spline, defined by a rather small number of grid nodes ($N_r \sim 20 - 50$),
typically spaced equally in $\log r$ over a range $r_{\rm max} / r_{\rm min} \gtrsim 10^6$; the truncation order of angular expansion $\ell_{\rm max}$ depends on the shape of the density profile -- with higher values capturing more sub-structure and asymmetries -- and it is usually $\lesssim 10$. Note that $| m | \leq \ell_{\rm max}$.

Our approach is as follows. First, we run a simulation with a fully responsive DM halo, as described above in Sec.~\ref{sec:bench} for $\sim2$~Gyr. Then, we extract the DM component -- that is, the phase space coordinates of the DM particles -- at regular time intervals $\delta t \approx 10$~Myr, starting from $t = 0$, and fit a potential of the form given by Eq.~\eqref{eq:phi_ana} to each of these snapshots. We ensure that the N-body distribution is well centred both in space (and velocity) by applying the `shrinking sphere method' \citep{pow03o}, as implemented in \pynbody. Note that we do not apply a correction to the halo's alignment as it is not expected to tumble, given the simple nature of the setup. However, note that the alignment of the halo, either of its spin or principal axes -- in addition to a correction for the halo's motion -- may be necessary in the case of a cosmological, or zoom-in, simulation \citep[see e.g.][]{low11a,san20b,ara25a}.

Crucially, \agama\ allows us to impose a specific symmetry onto the analytic potential (e.g.~spherical, axisymmetric, etc.); we choose not to assume any prior symmetry. This is particularly relevant for our future, similar experiments where we apply this framework to cosmological or zoom-in simulations, in which DM halos do not generally have any type of symmetry, especially if they are not settled \citep[e.g.][]{ara25a}.\footnote{Not imposing a symmetry has the additional benefit of making the reconstructed potential invariant with respect to the rotation of the coordinate axes (e.g.~when the halo tumbles).} An overview of relevant parameters and their values is provided in Tab.~\ref{tab:agama}.

\begin{table}
\begin{center}
\caption{\agama\ potential reconstruction parameters. Parameters not listed here take their default value (in version~1.0.151 of the library).
}
\label{tab:agama}
\begin{tabular}{lcl}
Internal name & Value &  Remarks \\
\hline
{\tt type} & Multipole & Potential type \\
{\tt symmetry} & none & Potential symmetry \\
{\tt rmin}  & 0.15 kpc & Minimum radius$^a$ \\
{\tt rmax}  & 150 kpc  & Maximum radius$^b$ \\
{\tt gridSizeR} & 25 & Number of radial grid points \\
{\tt lmax} & 6  & Maximum multipole order$^c$ \\
\hline
\end{tabular}
\end{center}
\begin{list}{}{}
\item Notes. $^a$We have checked that the choice of {\tt rmin} does not affect the reconstruction error close to the centre, as long as it is on the order of a few AMR cells. $^b$Chosen so as to fully enclose the DM halo. $^c$We also tried values 14 and 20 (see text for details).
\end{list}
\end{table}

Our standard choice for the truncation multipole order is $\ell_{\rm max} = 6$. However, we have also experimented with higher values $ \ell_{\rm max} \in \{14, 20\}$. We found that the results for $\ell_{\rm max} = 14$ and $\ell_{\rm max} = 20$ are very similar to $\ell_{\rm max} = 6$, and thus we will focus on the latter moving forward. However, we do show some results of the former for completeness, but defer its presentation to the Appendix (Sec.~\ref{app:lmax}).

\begin{figure}
\centering
\includegraphics[width=\columnwidth]{ 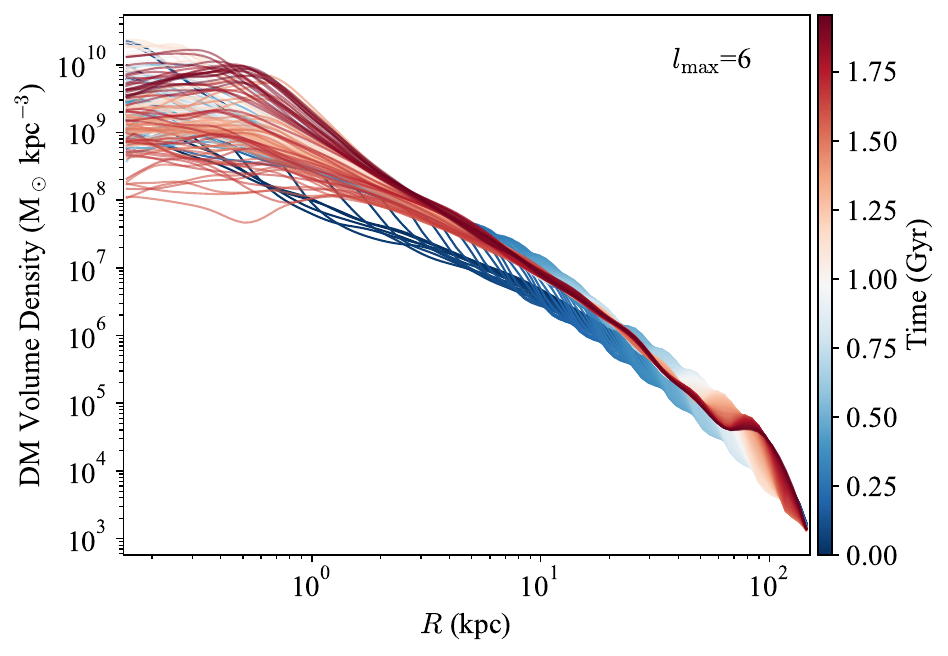 }
\includegraphics[width=\columnwidth]{ 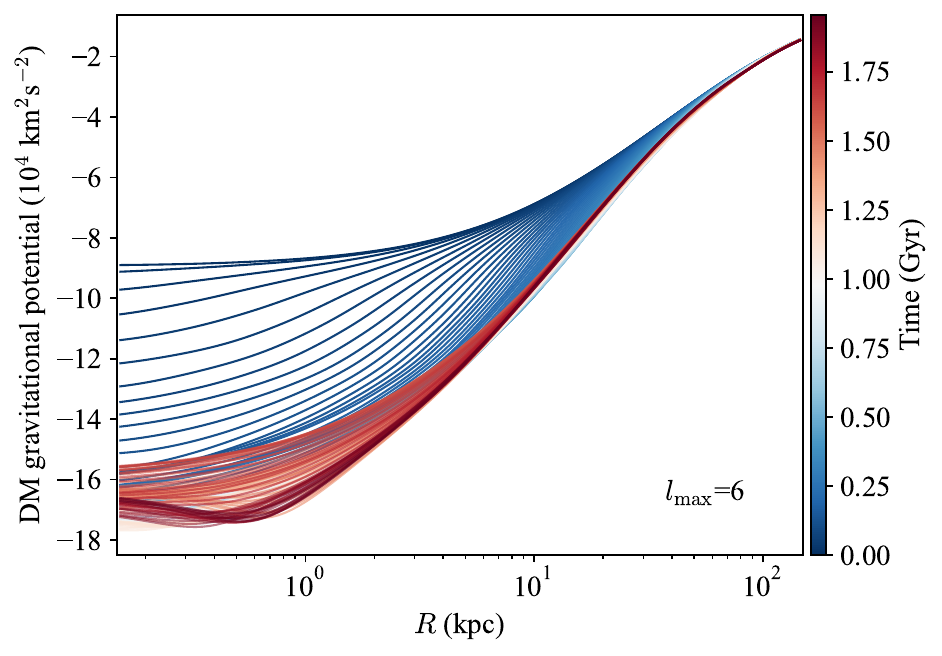 }
\caption{Time evolution of the reconstructed density (top) and reconstructed potential (bottom) of the DM component in the N-body responsive halo model, adopting $\ell_{\rm max} = 6$. The profiles for $\ell_{\rm max} = 14$ are similar (see Appendix \ref{app:lmax}).}
\label{fig:phi_ana}
\end{figure}

Fig.~\ref{fig:phi_ana} shows the evolution of the mass volume density (top) and potential (bottom) of the DM component in the `N-body responsive halo'  (i.e.~the benchmark) run. Note that: 1) these are {\em reconstructed} quantities, and 2) the plots show the profile along the $x$ coordinate in the mid-plane (i.e.~$z = 0$). Given the system's initial axisymmetry, we do not expect an azimuthally averaged profile to be significantly different.
Clearly, the DM density increases with time (top panel) and correspondingly its potential deepens but its radial gradient flattens (bottom panel) due to accumulation of baryons at the halo's centre.

\subsubsection{Reconstruction Verification} \label{sec:ver}

In order to evaluate the accuracy of the potential reconstruction (Eq.~\ref{eq:phi_ana}), we calculate an independent estimate for the potential induced by the distribution of DM particles using a tree-based method \citep{bar86b}, as implemented in the {\sc pytreegrav} module \citep{gru21a}.
The most relevant parameter in the latter is the `opening angle' $\theta \in (0,1]$, which controls the accuracy of the estimate, whereby smaller values yield a more accurate reconstruction, while simultaneously increasing the computational cost. Our standard choice is $\theta = 0.7$, but we also try smaller values (0.3 and 0.5). We regard the {\sc pytreegrav} potential as the `true' potential.\footnote{We note that this is the true {\em N-body} potential, not the underlying, smooth (but unknown) potential. The former is the one that particles actually feel. } To which extent this choice is justified is further explored in the Appendix (Sec.~\ref{app:theta}).

We quantitatively compare the \agama\ reconstruction with the true potential as follows. First, we generate a sample of $10^4$ random positions uniformly distributed within the volume of a sphere of radius equal to\footnote{Note this volume fully encloses the stellar disc in our benchmark simulation.} 20~kpc. Then, we evaluate the force\footnote{Vectors are denoted using a boldface font.} vector, $\mathbf{a}$, at each of these positions with each potential, and calculate the difference in magnitude and direction of $\mathbf{a}$ between the \agama\ reconstruction and the true potential, as a function of spherical radius $r$, and average the result over $r$.

\begin{figure}
\centering
\includegraphics[width=\columnwidth]{ 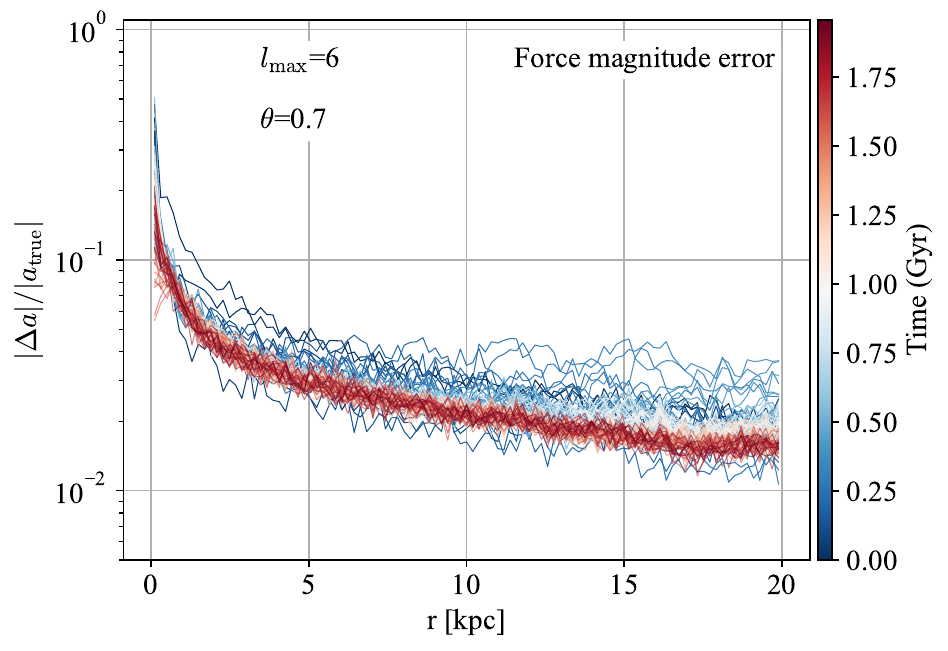 }
\includegraphics[width=\columnwidth]{ 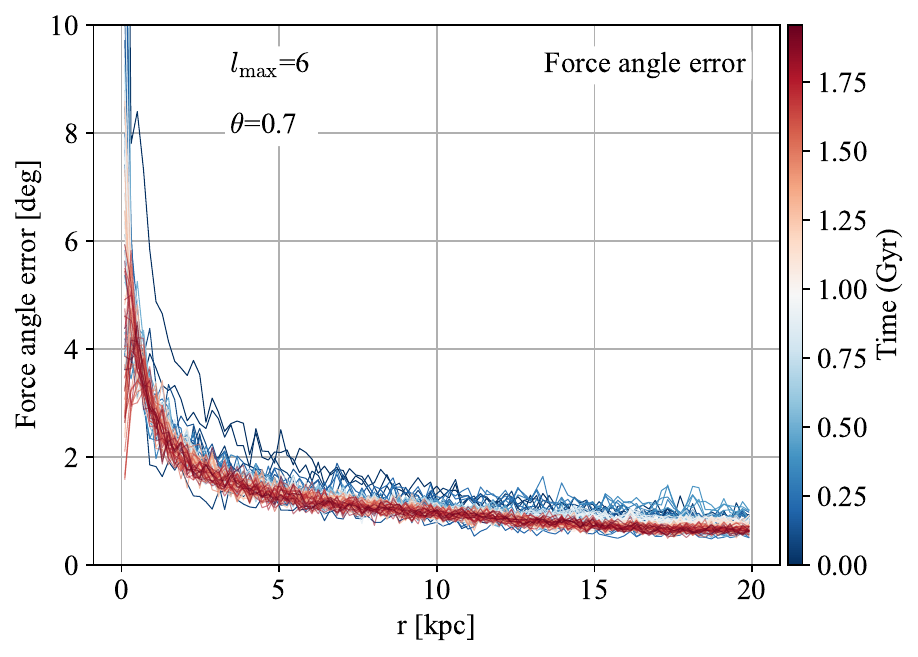 }
\caption[  ]{ Evaluating the potential reconstruction. Top: Relative force magnitude error. Bottom: Force directional error.
Note that an angle error of $\lesssim 10$~deg implies that the force vectors are aligned to within 2 percent. }
\label{fig:eval}
\end{figure}

Fig.~\ref{fig:eval} displays these quantities, corresponding to our standard choices $\ell_{\rm max} = 6$ and $\theta = 0.7$. The top panel shows the relative force magnitude error,
$
	{\left\vert \Delta \mathbf{a} \right\vert} / {| \mathbf{a}_{\rm true} |} \,,
$
where \mbox{ $\Delta \mathbf{a} \equiv \mathbf{a}_{\rm true} - \mathbf{a}_{\textnormal{\sc agama}}$}. The bottom panel shows the force misalignment, expressed as an angle $\alpha$ (in degree), where the angle is obtained from\footnote{This approach is more stable numerically compared to calculating the $\arccos \alpha$ from the dot product.}
$$
	\arctan \alpha = \frac{| \mathbf{a}_{\rm true} \times \mathbf{a}_{\textnormal{\sc agama}} |}{ \mathbf{a}_{\rm true} \cdot \mathbf{a}_{\textnormal{\sc agama} } } \,.
$$
In either panel, each curve corresponds to a given simulation epoch ($t$) with values between 0 and $\sim2$~Gyr, in steps of $\delta t \approx 10$~Myr, as indicated by its colour and the colour bar on the right.

This exercise indicates that the force magnitude error is limited to a few percent throughout much of the volume within \mbox{20~kpc},
except closet to the centre, where it rises to \mbox{$\lesssim 10$ percent}. Similarly, the force misalignment is of order a few degree over much of the volume, climbing up to \mbox{$\sim 10$~deg} at the centre.\footnote{A median angular error of 1~deg corresponds to a difference of $\sim1.7$~percent between the force directions. These estimates scale roughly linearly with angle up to $\sim 10$~deg.} These results are comparable to those found by others using different reconstruction methods and different simulations \citep[e.g.][]{aro24a}. Neither increasing $\ell_{\rm max}$ to 14 or even 20, nor estimating the tree (`true') potential with $\theta = 0.3$ or $\theta = 0.5$ results in a significant improvement of the reconstruction (at the expense of a higher computational cost).
Both these assertions are demonstrated by Fig.~\ref{fig:eval2}, displayed in the Appendix. We conclude that the potential reconstruction is converged with respect to $\ell_{\rm max}$.

These results, taken at face value, demonstrate that the \agama\ potential reconstruction reproduces the true potential with sufficient accuracy, thus verifying our potential reconstruction method. Its (ultimate) validation will come from the comparison of the disc that forms in simulations using this reconstruction to that of the benchmark run, discussed next.\footnote{We adhere to the definitions of `verification' and `validation' of \citet{cal02a}.}

\subsection{Surrogate Halo Simulation} \label{sec:surr}

The next step in our strategy is to run a simulation in which we replace the gravitational effect of the fully responsive N-body DM halo by its analytic reconstruction $\Phi_{\rm ana}$, a `surrogate halo'. This replacement can be accomplished with the standard \ramses\ code with a few modifications, complemented with some features provided by the \agama\ library, as described next.

\ramses\ implements the time evolution of an N-body~/~hydrodynamical composite system via solution of the Vlasov-Poisson equations,
\begin{align}
	\frac{{\rm d}\mathbf{r}}{{\rm d}t} & = \mathbf{v}(\mathbf{r}, t)  \label{eq:vpe1} \\
	\frac{{\rm d}\mathbf{v}}{{\rm d}t} & = -\nabla \Phi(\mathbf{r}, t)  \label{eq:vpe2}
\end{align}
where the total gravitation potential $\Phi$ on the AMR grid is given by \citep{gui11a}
\begin{equation} \label{eq:pot1}
	\nabla^2 \Phi(\mathbf{r},t) = 4\pi G \sum_i \rho_i(\mathbf{r},t) \, ,
\end{equation}
Here, $\rho_i(\mathbf{r},t)$ is the mass density of component $i$ at time $t$. The simulations described here and in Sec.~\ref{sec:bench} contain three components: dark matter (DM), gas, and stars.\footnote{Note that, while the DM component and the gas component are present in the simulation from the outset, stars form thereafter from the gas. } Thus, in our setup,\footnote{We omit the argument from now on, with the understanding that the potential depends on the position and time.}
\begin{equation} \label{eq:pot2}
	\nabla^2 \Phi= 4\pi G \left( \rho_{\textnormal{\sc dm}} + \rho_{\rm gas} + \rho_{\rm star} \right) \, .
\end{equation}
In a fully self-consistent simulation, each component therefore makes a contribution to the gravitational potential and affects, and is affected by, the gravitational potential of the other components. 

In our benchmark simulation, $\rho_{\textnormal{\sc dm}}$ in Eq.~\eqref{eq:pot2} is provided by the distribution of $N_{\textnormal{\sc dm}}$ collisionless particles of equal mass, $m_{\textnormal{\sc dm}}$. For our test simulation, we need to replace it by the density that self-consistently (via Poisson's equation) relates to $\Phi_{\rm ana}$; the required density can be readily calculated with \agama. Schematically,
$$
	\Phi_{\rm ana} \stackrel{\rm AGAMA}{\to} \rho_{\rm ana}  \stackrel{{\textnormal{\sc Ramses}}}{\to} \Phi  \stackrel{{\textnormal{\sc Ramses}}}{\to} \nabla \Phi
$$
It is instructive to note that all of these steps are accomplished within \ramses.\footnote{To this end, we make use of \agama{}'s Fortran interface, which can be readily integrated into the \ramses\ source code. } 

This approach is not without shortcomings. The most crucial one is that computation of the density involves second derivatives of the potential, which are susceptible to numerical noise. In addition, the density may become infinite at the coordinate's origin, owing to the nature of the expansion's coefficients.
We have never encountered this issue in our simulations, likely due to the negligible chance of a particle being at (or close enough to) the origin of the coordinate system. But an occasional issue is the appearance of negative densities. These appear exclusively in very low-density regions, and are caused by small-amplitude oscillations of the reconstruction around zero. While these are not an issue in general, they do cause a runtime issue in \ramses.\footnote{The code does check for negative densities in general, but not for user-provided functions; this responsibility falls onto the user.} In order to avoid this, we introduce a smooth `clipping' function,
$$
	\rho \leftarrow \frac{1}{2} \left( \rho + \sqrt{\rho^2 + \rho^2_{\rm min} } \right)  
$$
that is applied whenever $\rho < 0$, and which reassigns the density value such that $\rho \geq \rho_{\rm min}$ at all times. The parameter $\rho_{\rm min}$ is the minimum acceptable value for the density on the grid, and it is an intrinsic parameter of \ramses, with a default value of $10^{-10}~n_{\rm cu}$ (in code units $n_{\rm cu} \sim 10^{-3}$~\pcc), which we adopt in our experiments. In keeping with \ramses's nomenclature, we refer to this approach as {\tt rho\_ana}.

Given all these limitations, a better approach would be to make use of $\Phi_{\rm ana}$ directly when computing the potential in \ramses; alas, we are yet to implement this feature in the code. But \ramses\ does offer an alternative way to use an analytic gravitational field. Rather than inserting an additional density source term in Eq.~\eqref{eq:pot2}, one can directly specify its corresponding acceleration, and add it directly to the right-hand side of Eq.~\eqref{eq:vpe2}; as is the case of the density, the required acceleration can be readily calculated with \agama. Schematically,
$$
	\Phi_{\rm ana} \stackrel{\rm AGAMA}{\to} \mathbf{g}_{\,\rm ana}  \stackrel{{\textnormal{\sc Ramses}}}{\to} \nabla \Phi
$$
We refer to this approach as {\tt grav\_ana}. It is worth noting that the computation of the acceleration involves a single derivative and its therefore subject to less numerical noise, compared to the computation of the density. It also requires one step less than the {\tt rho\_ana} approach to arrive at $\nabla \Phi$. Therefore, it is expected to be overall computationally cheaper. Needless to say, we will explore and contrast both these approaches in our numerical experiment.\\

With the methodology now established, we proceed to set up and run a series of simulations, each designed to test a specific feature of our approach. Specifically, we consider the following experiments, including our benchmark and surrogate runs (q.v. Tab.~\ref{tab:runs}):

\begin{enumerate}
	\item[\bf N-body responsive halo (benchmark):] A self-consistent, N-body $+$ hydrodynamical simulation where the DM halo is fully responsive. This constitutes our benchmark simulation (described in detail in Sec.~\ref{sec:bench})
	\item[\bf N-body static halo:] Similar to the benchmark run, but where the DM particle positions are frozen at $t = 0$ over the full simulation timespan. Note the DM halo is {\em not} replaced by an analytic model.
	\item[\bf Analytic static halo:] The N-body model of the DM halo in the benchmark run is replaced by $\Phi_{\rm ana}(\mathbf{r}, t\equiv0)$.  In other words, the DM halo's potential is replaced by an analytic reconstruction fitted {\em and fixed} at $t = 0$, for the entire evolution of the system. The corresponding DM density is estimated via the {\tt rho\_ana} approach.
	\item[\bf Analytic evolving halo~/~{\tt rho\_ana}:] The N-body model of the DM halo in the benchmark run is replaced by an analytic reconstruction $\Phi_{\rm ana}(\mathbf{r}, t)$ fitted at different time steps $\delta t \approx 10$~\Myr, starting at \mbox{$t = 0$}. The density is estimated via the {\tt rho\_ana} approach. During the system's evolution, the closest analytic reconstruction in time for a given $t$ is used.\footnote{\citet{san20b}  find that linearly interpolating between the two closest time steps does not yield noticeably different results, provided $\delta t$ is `small' enough; in our case, $\delta t \approx 10$~Myr is a cadence high enough to make interpolation unnecessary. }
	\item[\bf Analytic evolving halo~/~{\tt grav\_ana}:] Similar to `analytic, evolving halo ~/~{\tt rho\_ana}', but where the analytic potential is used to directly calculate the force induced by the DM mass distribution, added directly to the right-hand-side of Eq.~\eqref{eq:vpe2}, i.e. the {\tt grav\_ana} approach.
	\item[] Each of these last two simulations represents a replay with a genuine `surrogate halo'.
\end{enumerate}

\begin{table*}
\begin{center}
\caption{ Simulation summary. Note that the only difference between the simulation setups is the nature of the DM halo potential. All quantities correspond to the disc. Columns are as follows: (1) DM halo descriptor; (2) Stellar mass; (3,4) Scalelength and central density; (5,6,7) Root-mean-square height, radial velocity dispersion, and vertical velocity dispersion of the stars; (8,9) gas mass and its scalelength.
}
\label{tab:runs}
\begin{tabular}{lcrrccccr}
DM halo model & $M_{\rm star}$ & $R_{\rm d, star}$$^a$ & $\Sigma_{\rm star}(0)$$^a$ & $z_{\rm RMS}$$^b$ & $\sigma_{R}$$^b$ & $\sigma_{z}$$^b$ & $M_{\rm gas}$ & $R_{\rm d, gas}$$^a$ \\
& ($10^9$~\Msun) & (kpc) & ($10^6$~\Msun) & (kpc) & (\kms) & (\kms) & ($10^{9}$~\Msun) & (kpc) \\
\hline
N-body, responsive (benchmark) & 2.324 & 3.33$\pm$0.10 & 31.4$\pm$2.93 & 0.64 & 37.66 &19.41 & 10.87 & 9.51$\pm$0.47 \\
N-body, static & 1.884 & 6.47$\pm$0.23 & 8.75$\pm$0.50 & 1.73 & 23.96 & 20.59 & 9.10 & 14.36$\pm$1.08 \\
Analytic, evolving (surrogate) & 2.241 & 3.22$\pm$0.08 & 33.0$\pm$2.70 & 0.68 & 35.11 & 21.77 & 11.18 & 10.45$\pm$0.80 \\
Analytic, static& 1.787 & 4.37$\pm$0.11 & 15.1$\pm$0.92 & 1.06 & 22.77 & 17.52 & 9.205 & 12.76$\pm$0.97 \\
\hline
\end{tabular}
\end{center}
\begin{list}{}{}
\item Notes. All quantities measured at $t \approx 2$~Gyr. $^a$~The error corresponds to the fit uncertainty, $^b$~Measured at $R = 2.2~R_{\rm d}$. 
\end{list}
\end{table*}

By design, each simulation differs from the previous one in a single aspect. Thus, the full suite allows for a differential analysis of the effect or stepwise replacing the DM halo's potential by both a frozen halo or an analytic (static/evolving) reconstruction. For instance, the benchmark run and the `N-body static halo' run are ideal to compare the effect of a non-evolving potential on the evolution of the disc, while maintaining the (initial) 'granularity' of the halo. Next, the runs `N-body static halo' and `analytic static halo' are designed to compare the effect of the DM halo's granularity on the evolution of the disc within a non-evolving potential. These two comparisons are relevant as it is not uncommon in the literature to use either static and / or analytic potentials to study the structural, dynamical, and kinematical properties of stellar discs \citep[e.g.][]{don13a,nob24a,str26a}.

The `analytic static halo' and `analytic, evolving halo' runs make it possible to compare the effect on the disc of a static potential compared to an evolving potential, where both are described by an analytic reconstruction. Notably, the use of an {\em analytic} reconstruction removes the noise inherent to the discrete nature of an N-body halo. The `analytic, evolving halo~/~{\tt rho\_ana}' and 'analytic, evolving halo~/~{\tt grav\_ana}' runs allow to compare the different approaches to `feed' the analytic potential to \ramses. While we do not expect a major difference in terms of physical results between these, we do anticipate that the former is computationally more expensive. The most relevant comparison is between these last two runs and the benchmark run, discussed next.

Of particular note is that all the simulations are run using the same exact \ramses\ setup described in see Sec.~\ref{sec:bench}, i.e. boxsize, min~/~max levels, etc. In particular, the same refinement criteria are applied throughout. We will return to this point in Sec.~\ref{sec:concl}.

\section{Validation} \label{sec:val}

\begin{figure*}
\centering
\includegraphics[draft=false, width=0.8\textwidth]{ 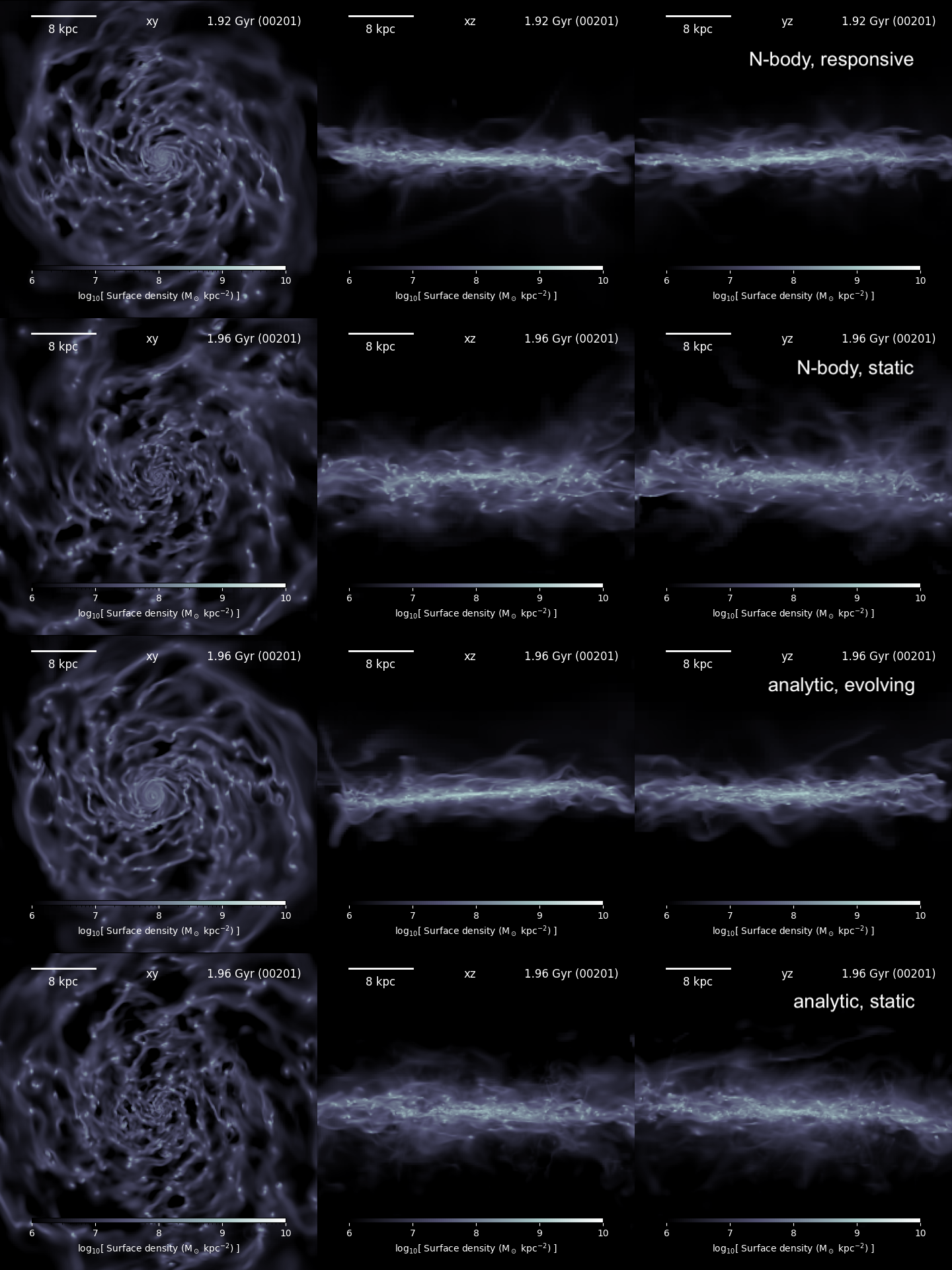 }
\caption[  ]{
State of the gaseous disc component after $\sim2$~Gyr of evolution along three orthogonal projections: $xy$ (left); $xz$ (centre); $yz$ (right). Each row corresponds to a different run. From top to bottom: N-body responsive halo (benchmark); N-body static halo; Analytic, evolving halo (surrogate); Analytic static halo. See Tab.~\ref{tab:runs}.
}
\label{fig:disc_gas}
\end{figure*}

\begin{figure*}
\centering
\includegraphics[draft=false, width=0.8\textwidth]{ 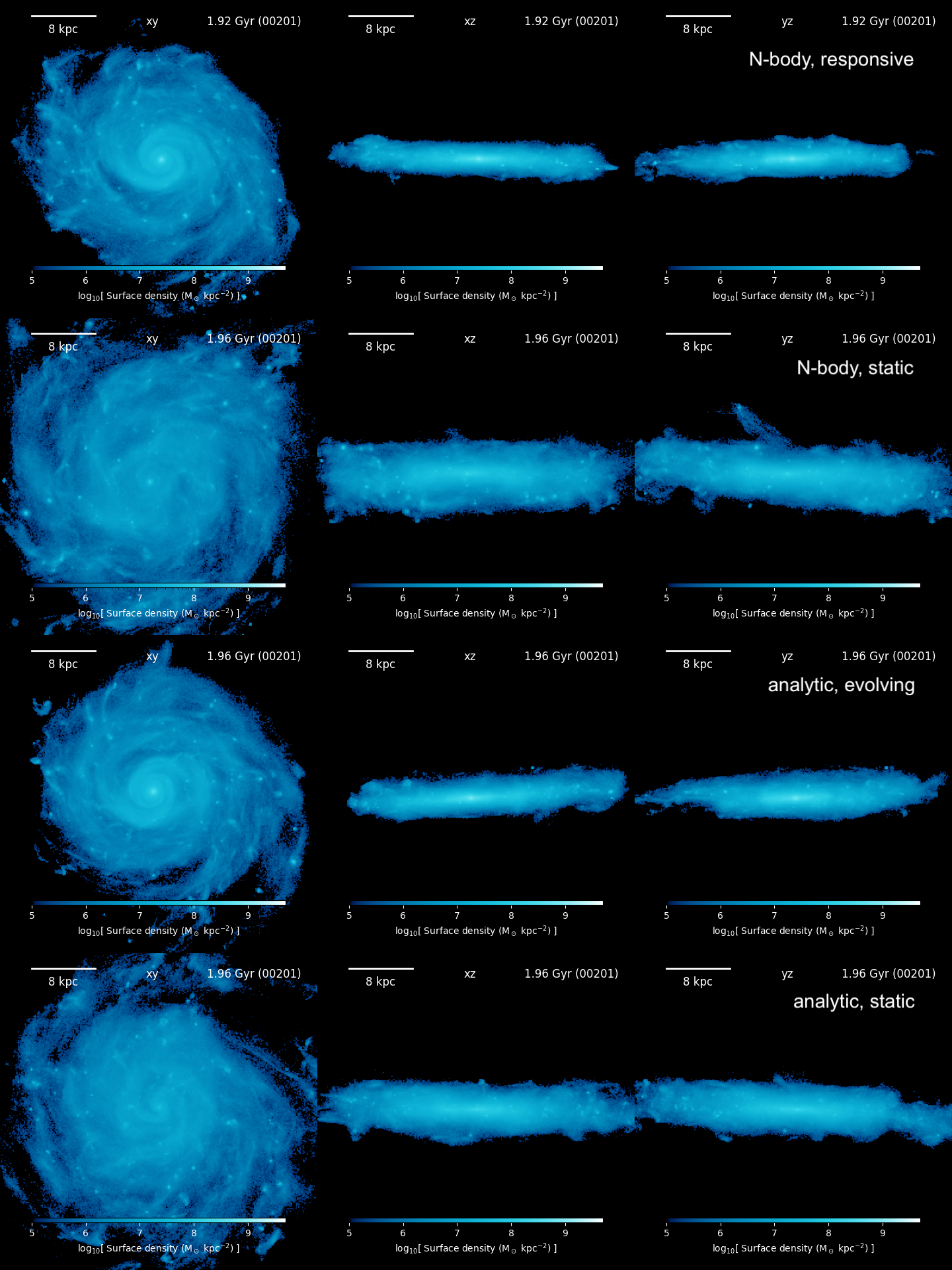 }
\caption[  ]{
State of the stellar disc component after $\sim2$~Gyr of evolution along three orthogonal projections: $xy$ (left); $xz$ (centre); $yz$ (right). Each row corresponds to a different run. rom top to bottom: N-body responsive halo (benchmark); N-body static halo; Analytic, evolving halo (surrogate); Analytic static halo. See Tab.~\ref{tab:runs}.
}
\label{fig:disc_stars}
\end{figure*}

The comparison between the benchmark run and the other runs is concerned with the disc -- both the gaseous and the stellar components -- that forms within the halo. Specifically, the analysis focuses on the disc's mass assembly, and its structural- and kinematic properties. We consider the mass assembly over the full evolution of the system, but the structural and kinematic properties are only compared at the final time step, i.e. after $\sim2$~Gyr of evolution. As an illustration, we present the state of the disc density, both of the gaseous component and the stellar component, after $\sim2$~Gyr of evolution in Figs.~\ref{fig:disc_gas} and \ref{fig:disc_stars}, respectively. Some relevant observables, discussed next, are summarised in Tab.~\ref{tab:runs}.

Upon completing the analysis, we find -- as anticipated above -- that the physical results from both `analytic, evolving halo' runs (i.e. {\tt rho\_ana} and {\tt grav\_ana}) are very similar (we also confirmed our expectation that the former is computationally more expensive to run). To avoid redundancy, in the following we omit results corresponding to the `analytic, evolving halo~/~{\tt rho\_ana}' run, and now refer to `analytic, evolving halo~/~{\tt grav\_ana}' simply as `surrogate halo'.

We draw attention to the fact that in all the figures that follow (Figs.~\ref{fig:sfh} - \ref{fig:psd}),
dot-dashed line-styles correspond to simulations featuring an {\em N-body} DM halo model; solid line-styles, to simulations featuring an {\em analytic} DM halo reconstruction. Thick line-styles correspond to simulations featuring an {\em evolving} halo; thin line-styles, to simulations featuring a {\em static} halo. The results corresponding to the benchmark run are highlighted by a red, thick, dot-dashed line-style throughout.

\subsection{Disc mass assembly} \label{sec:sfh}

\begin{figure}
\centering
\includegraphics[width=\columnwidth]{ 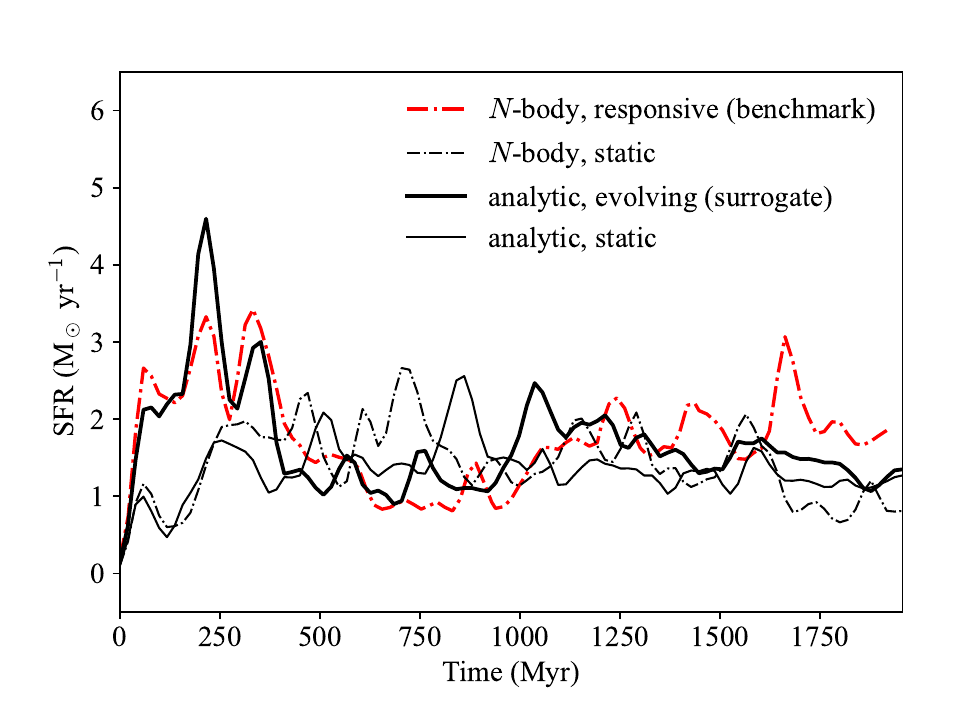 }
\includegraphics[width=\columnwidth]{ 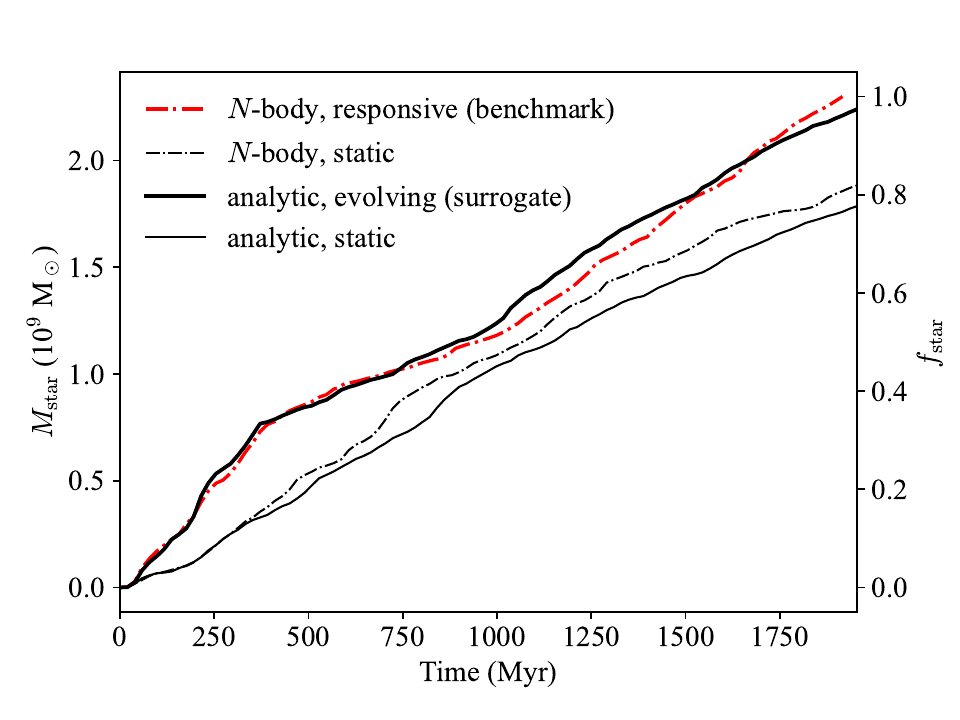 }
\includegraphics[width=\columnwidth]{ 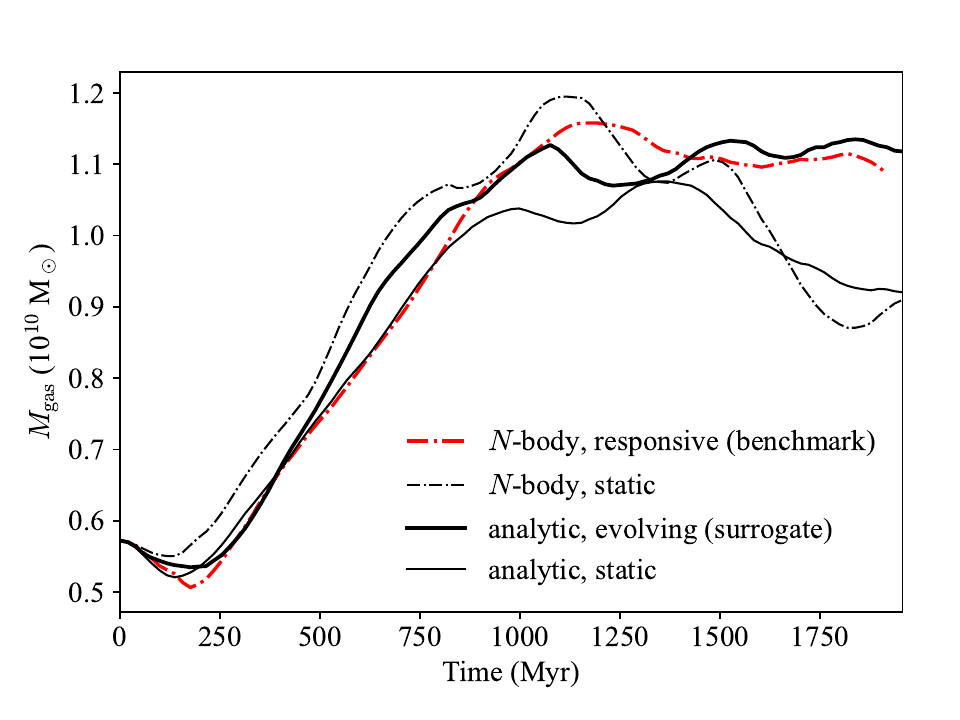 }
\caption[  ]{ Disc mass assembly. Top: Star formation history. For visualisation purposes, the SFH has been slightly de-noised by convolving the raw data with a Gaussian window of size $\sigma = 1$~sample. Middle: Stellar mass build-up as a function of time. The alternative $y$-axis indicates the stellar mass fraction of each run relative to the benchmark run at the last time step. Bottom: gas mass build-up within the disc as a function of time. Note the different unit mass scale in the middle and bottom panels. }
\label{fig:sfh}
\end{figure}

The bulk (or low-order) properties of the system, i.e. the star formation history (SFH) and the disc mass assembly, are inherently coupled, and are therefore analysed in tandem. Fig.~\ref{fig:sfh} displays the evolution of these observables for all simulations.

The top panel displays the SFH. The star formation rate is naturally stochastic,\footnote{For visualisation purposes, the SFH has been slightly de-noised by convolving the raw data with a Gaussian kernel of size $\sigma = 1$~sample.} regulated by the interplay between gas cooling and heating (via stellar feedback), with a mean level on the order of \mbox{$1 - 2$~\Msun~\pyr} over the full time span irrespective of the run. The benchmark run has initially a level of star formation, on average, closely matched by the `surrogate halo' run, which is significantly higher compared to the static halo runs. These match the level observed in the other runs only during $t \approx 500 - 1000$~Myr. The difference could be due to the presence of a DM overdensity arising from the halo's contraction, which is naturally present in a responsive halo and to some extent captured in its analytic reconstruction, but absent in a static halo, by design. Such an overdensity could deepen the central potential and enhance the confinement and compression of the gas, potentially leading to the elevated star-formation activity observed in the responsive halo model. Conversely, its absence leads to a lower level of star formation activity. The difference could in part be caused by different levels of spatial resolution between the simulations. We address this point in Sec.~\ref{sec:cav}.

The differences in the SFH between the benchmark run and the other models become more apparent when comparing the stellar mass formed over a specific period of time, as shown in the middle panel of Fig.~\ref{fig:sfh}. The disc mass in the benchmark run amounts to $M_{\rm d} \approx 2.3 \times 10^9$~\Msun\ by $\sim 2$~Gyr. Note that this corresponds to roughly 10 percent of the initial baryonic mass in the system.\footnote{The {\em total} (i.e.~gas and stars together) disc mass is roughly 50~percent of the hot gas halo mass, and thus lower than the value predicted by the \citet{mo98a} formalism; likely, the disc would become more massive if the system continued evolving. We defer an in-depth analysis of this circumstance to future work.} The stellar mass build-up in the surrogate halo run reproduces that in the benchmark run. It is interesting that, despite the stochastic nature of the star-formation process, reflected in the noticeably different SFH between the benchmark run and the surrogate halo run, the stellar mass assembly proceeds more or less along the same locus.
In contrast, both the disc in each of the static halo runs ends up with a lower, total stellar mass, which by $\sim 2$~Gyr corresponds to roughly 80~percent of the stellar mass in the benchmark run; the smallest total stellar mass forms within the `analytic, static halo' run. In either case, the cause of the mass difference is the large difference in the individual SFRs over the first $\sim 500$~Myr, as discussed above.

While the mass assembly of the stellar disc is naturally monotonic (roughly linear) in time, the mass build-up of the gas disc displays a rather different behaviour, which is qualitatively similar across all models. During the first $\sim250$~Myr of evolution, the gas fraction within the disc region\footnote{The `disc region' is defined here as the cylinder around the system's centre of mass with a radius of 20~kpc, and an equal height. The latter is deliberately chosen to ensure the disc is fully contained within. Smaller values do not affect the result, as the disc (both its gaseous and stellar component) are concentrated around the mid-plane.} drops due to the conversion of gas into stars, but also -- albeit to a lesser extent -- due to the outflow of gas as a result of stellar feedback. Thereafter, accretion from the surrounding hot gas via cooling, enhanced by the enriched material injected by stellar activity, leads to a progressive gas mass build-up in the disc for $\sim1$~Gyr. At this point, the models show differing behaviours. In the benchmark run, the gas mass build-up flattens, suggesting that the star-formation process (i.e.~accretion of gas $\to$ formation of stars $\to$ outflows due to stellar feedback) has become {\em self-regulated}, as is commonly found in cosmological- and standalone simulations \citep[e.g.][respectively]{sch10a,nob24a}.

The build-up of the gas disc in the surrogate halo run proceeds in a similar fashion to the benchmark run, featuring the three distinct phases: 1) initial drop; 2) monotonic (linear) increase; and 3) self-regulation. This self-regulation may be the reason why the stellar mass assembly proceeds in a similar way across these two models, as discussed above. The gas disc growth in the static halo runs proceeds along the first two phases too, but it does not show a self-regulated phase as clearly. Why these models behave this way we can only speculate. The relevant point is that they do not recover the behaviour of the gas disc growth in the benchmark run, while the surrogate halo run does.

\subsection{Structure and kinematics of the stellar disc} \label{sec:disc_star}

Beyond bulk (or low-order) properties of the system, we are interested in the high-order properties of the stellar disc that forms within. To this end, we focus on the following structural and kinematical observables. In the first category: the stellar surface density ($\Sigma_{\rm star}$) and the root-mean-square (RMS) of the height of stars above the mid-plane ($z_{\rm RMS}$). In the second category: the rotational velocity around the centre ($v_\phi$), the radial velocity dispersion ($\sigma_R$), and the vertical velocity dispersion ($\sigma_z$). The analysis of the gaseous disc's observables is presented in Sec.~\ref{sec:disc_gas_struct}.

\subsubsection{Surface density and vertical profile} \label{sec:disc_star_struct}

Fig.~\ref{fig:disc_star_struct} displays the stellar surface density (top) and the RMS height of stars above the mid-plane (bottom) for the benchmark run (thick, red, dot-dashed curve) and for the other runs (black curves of different styles) at the end of the simulation. Both quantities correspond to azimuthal averages as a function of cylindrical radius $R$. The red, dashed line corresponds to an exponential profile with scalelength $R_{\rm d} \approx 3.3$~kpc, fitted\footnote{Vertically displaced by hand for visual clarity.} to the thick, red, dot-dashed curve in the range $R \in [ 1, 17.5 ]$~kpc. The surface density profile of disc in the benchmark run thus features an exponential profile, consistent with observations and theory \citep{pat40a,vau59a,fre70a,elm13a,wu20a}, as well as consistent with the assumption underlying \citeauthor{mo98a}'s formalism.

The surface density profile of the disc in all the other runs follows an exponential profile too, but with significantly different scalelength $R_{\rm d}$ and central surface density $\Sigma_{\rm star}(0)$. In the static halo runs, the disc profile is shallower, and more extended, compared to the disc in the benchmark run. In the surrogate halo run the disc follows a profile that is similar to that in the benchmark run. Fitting an exponential profile to each of the disc profiles over the range $R \in [ 1, 17.5 ]$~kpc confirms these findings (cf. Tab.~\ref{tab:runs}, column 3). Remarkably, the parameters of the surface density profile in the surrogate halo agree with those of the disc in the benchmark run, within their respective uncertainty.

The individual vertical profiles are shown in the bottom panel of Fig.~\ref{fig:disc_star_struct}. To facilitate a quantitative comparison, we have measured the $z_{\rm RMS}$ for each disc at $R = 2.2~R_{\rm d}$, which is the radius where a self-gravitating exponential disc reaches its peak circular velocity, and is commonly adopted for kinematic and dynamical measurements \citep[][]{fuj18a,bla23a,che25a}; these values are reported in Tab.~\ref{tab:runs} (column 4). In addition, we provide edge-on projections of each disc's density in configuration space (Fig.~\ref{fig:disc_stars}, central and right columns). 

The disc in the surrogate halo run has a comparable thickness to the disc in the benchmark run at radii $R \lesssim 10$~kpc, roughly equivalent to 3$R_{\rm d}$, and appears thicker at larger radii. The profiles of the discs in the static halo runs are thicker across all radii, and flare substantially, indicating significant vertical  heating. We believe that the reason for the thinner disc in the benchmark run compared to the the disc in the other discs is the presence of a higher concentration of DM towards the mid-plane as a result of the DM halo's contraction accompanying the collapse of the gas \citep[][]{blu86a,gne04a,sel05a}.
Such an oblate, inner structure increases the potential, and thus the restoring force, close to the mid-plane, resulting in smaller vertical orbit amplitudes for a fixed vertical orbit energy. Indeed, if we assume that to first order $z_{\rm RMS} \propto \sigma^2_z / \Sigma_{\rm tot}$ (self-gravitating isothermal sheet), then a higher $\Sigma_{\rm tot}$ naturally leads to a smaller $z_{\rm RMS}$ for a fixed $\sigma_z$. The agreement between the vertical profile of the disc in the surrogate halo run and in the benchmark run suggests that the time-dependent potential reconstruction captures this feature with enough accuracy in the inner galaxy ($R \lesssim3~R_{\rm d}$). At larger radii, the difference could also be in part due to the disc not being fully settled yet (cf. Fig.~\ref{fig:dform}).

\subsubsection{Kinematics} \label{sec:disc_star_kin}

The three panels in Fig.~\ref{fig:disc_star_kin} display the average rotational velocity of the stars (top), their average radial velocity dispersion (middle), and their average vertical velocity dispersion (bottom), for each of the runs. In line with previous figures, the result corresponding to the benchmark run is shown by a thick, red, dot-dashed curve, while the results for the test runs are indicated by black curves of different styles. We report the value of both the velocity dispersions for each disc at $R = 2.2~R_{\rm d}$ in Tab.~\ref{tab:runs} (columns~5 and 6).

\begin{figure}
\centering
\includegraphics[width=\columnwidth]{ 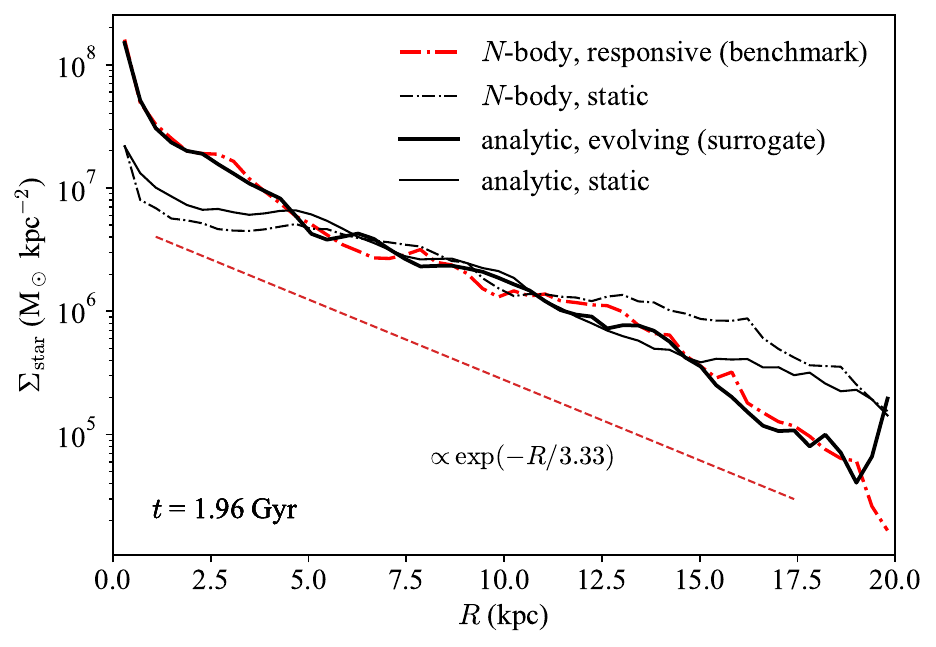 }
\includegraphics[width=\columnwidth]{ 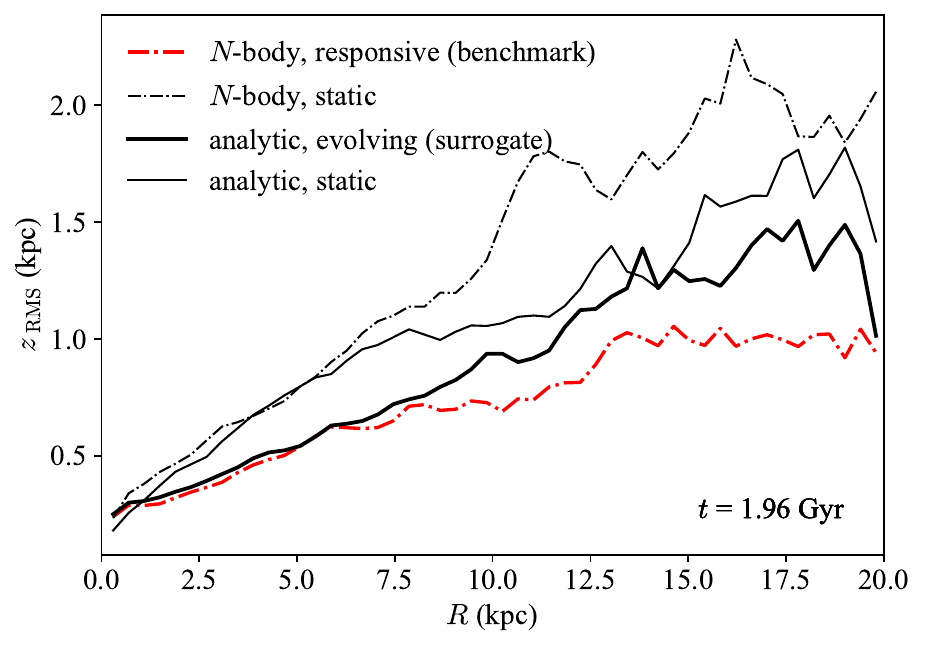 }
\caption[  ]{ Structural properties of the stellar disc after $\sim 2$~Gyr of evolution. Top: Surface density. The red, dashed line corresponds to an exponential profile with scalelength $R_{\rm d} \approx 3.3$~kpc, fitted to the thick, red, dot-dashed curve in the range $R \in [ 1, 17.5 ]$~kpc, but vertically displaced for visual clarity. Bottom: Root-mean-square of the height $z$ of stars above the mid-plane. Both quantities correspond to azimuthal averages as a function of cylindrical radius $R$.}
\label{fig:disc_star_struct}
\end{figure}

\begin{figure}
\centering
\includegraphics[width=\columnwidth]{ 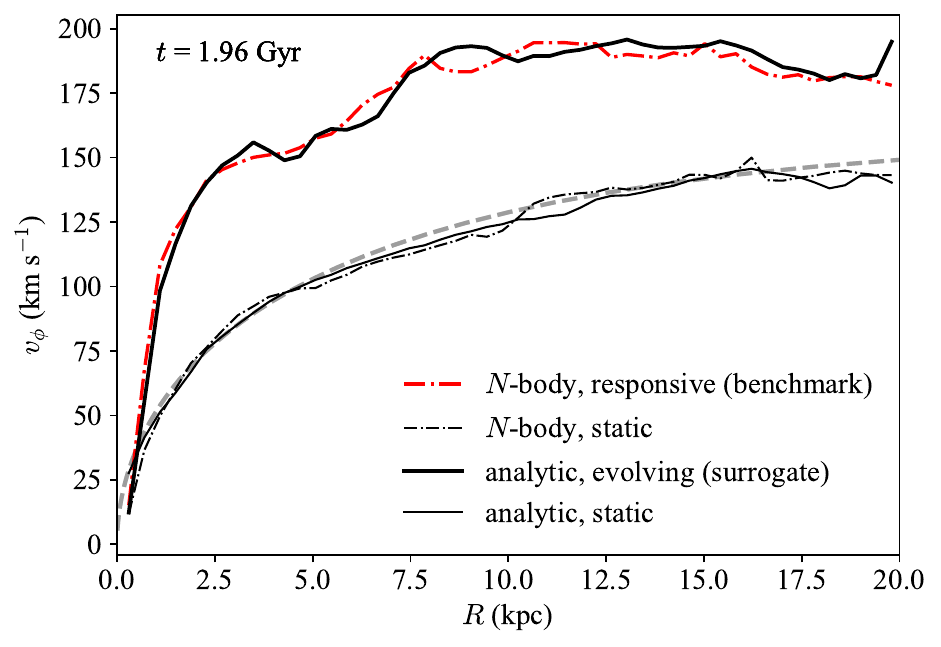 }
\includegraphics[width=\columnwidth]{ 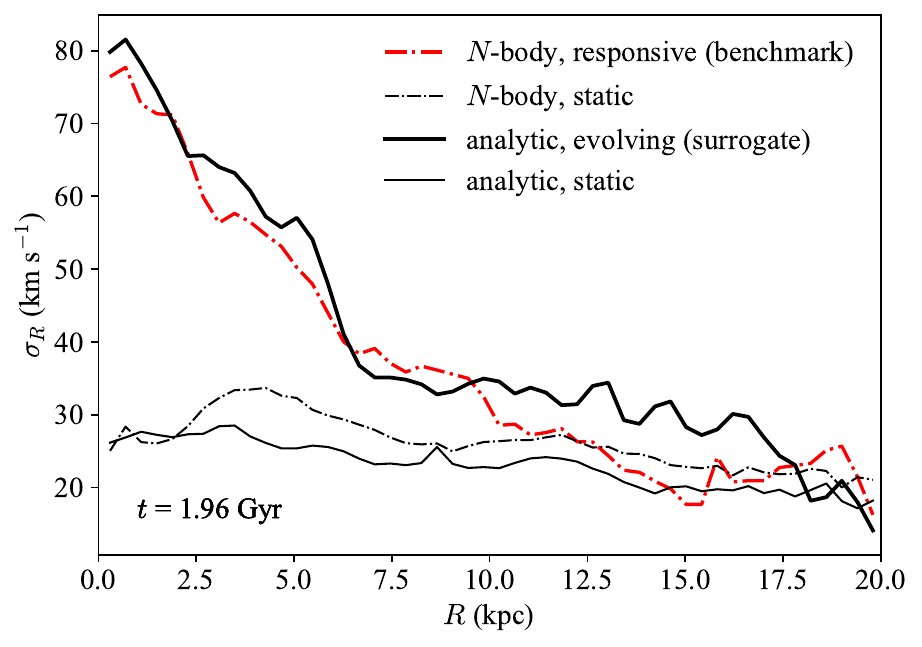 }
\includegraphics[width=\columnwidth]{ 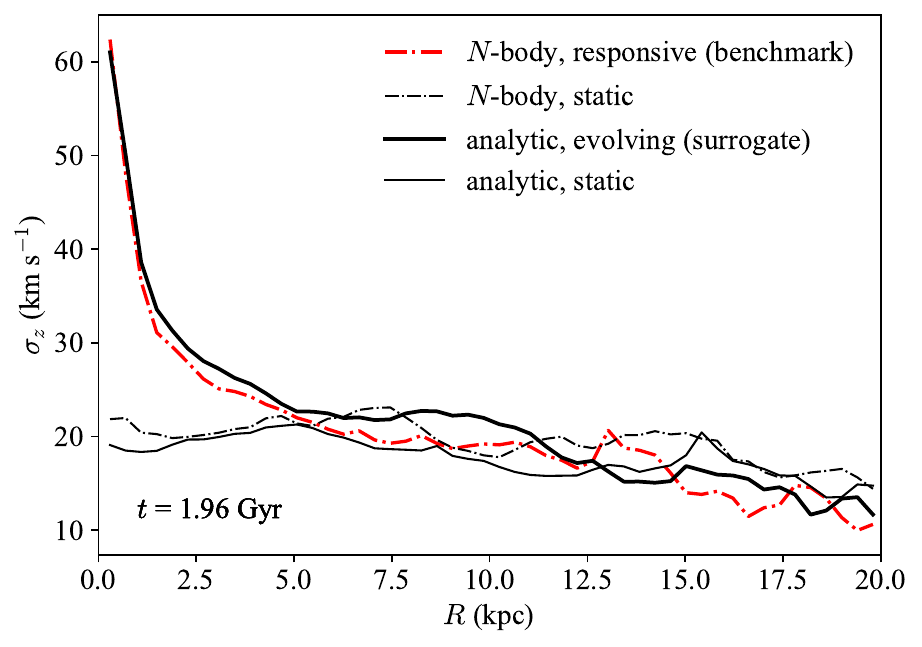 }
\caption[  ]{ Kinematic properties of the stars after $\sim 2$~Gyr of evolution. Top: Rotational velocity around the galaxy. The rotation velocity induced by the {\em initial} DM potential is shown by the thick, grey, dashed curve. Middle: Radial velocity dispersion. Bottom: Vertical velocity dispersion. All quantities correspond to azimuthal averages as a function of cylindrical radius $R$. }
\label{fig:disc_star_kin}
\end{figure}

Similar to the results previously discussed, the runs with a static halo (both N-body and analytic) differ significantly from the benchmark run, while the run with an surrogate halo broadly agrees with the latter. The rotation velocity of the stars around the galaxy in the static halo runs shows a significant difference to that in the benchmark run. This difference can be understood as a consequence of two circumstances: 1) that the DM halo is the dominant component of the circular speed in the galaxy; and 2) the evolution (or non-evolution) of the DM potential across different runs. Specifically, in the runs with an evolving halo (benchmark and surrogate), the potential well deepens considerably with time (cf. Fig.~\ref{fig:phi_ana}, bottom panel),
which naturally leads to an increase in the amplitude of the circular speed, and in consequence, of the mean rotational velocity of the stars around the galaxy \citep[cf.][]{mo98a}. In the static halo runs, the DM halo potential, by definition, does not evolve. Therefore, the final rotation velocity of the stars reflects the initial profile, corresponding to a {\em shallower} DM potential (shown by the thick, grey, dashed curve). Based on Fig.~\ref{fig:phi_ana}, we may provide a crude consistency check. The figure shows that the potential becomes a factor $\sim2.2$ stronger at (and around) the centre, implying that the circular speed curve increases by a factor $\sqrt{2.2} \approx 1.5$, in rough agreement with the difference in the amplitude between the `evolving halo' runs and the static halo runs displayed in the top panel of Fig.~\ref{fig:disc_star_struct}. The resulting rotational velocity of the disc in the benchmark run demonstrates that the DM halo contracts significantly as a result of the condensation of baryons towards the centre as the disc assembling proceeds.

The comparison of the individual velocity dispersion profiles shown in the middle and bottom panels of Fig.~\ref{fig:disc_star_kin} reveals, as we have comparisons of other observables, that the disc in the surrogate halo run broadly agrees with the disc in the benchmark run. This is true for both the radial velocity dispersion and the vertical velocity dispersion, although there is a noticeable difference of order 10~\kms\ in $\sigma_R$ at $R \gtrsim 10$~\kms\ (or $\gtrsim3~R_{\rm d}$) between the discs in these runs. We believe this difference, apparent over the same range as the difference in $z_{\rm RMS}$, is not of concern, as it relates to the outer, very low-density regions of the disc, which may well still be in the process of dynamical settling (cf.~Fig.~\ref{fig:dform}). The disc evolving within a static DM halo (represented either by a N-body model or by an analytic reconstruction) displays a roughly flat velocity dispersion profile in both $R$ and $z$, at all radii and in particular towards to the galaxy's centre, which seems unphysical and is observed neither in external galaxies \citep[q.v.][]{van11d} nor in the Milky Way \citep[q.v.][]{hun25a}.

The agreement of the velocity dispersion profiles of the disc that forms in the surrogate halo with that in the benchmark run is quite remarkable, in particular in the vertical direction; this lends further support to our belief that the difference in  $z_{\rm RMS}$ discussed above is due to a reduced restoring force in the surrogate halo run due to the (partial) absence of a `dark disc'. It may be that, for this same reason, the $z_{\rm RMS}$ profile of the disc in each of the static halo runs has a relatively large amplitude across all radii, despite not being kinematically hot in the vertical direction (low $\sigma_z$).

\subsection{Structure and kinematics of the gaseous disc} \label{sec:disc_gas_struct}

\begin{figure}
\centering
\includegraphics[width=\columnwidth]{ 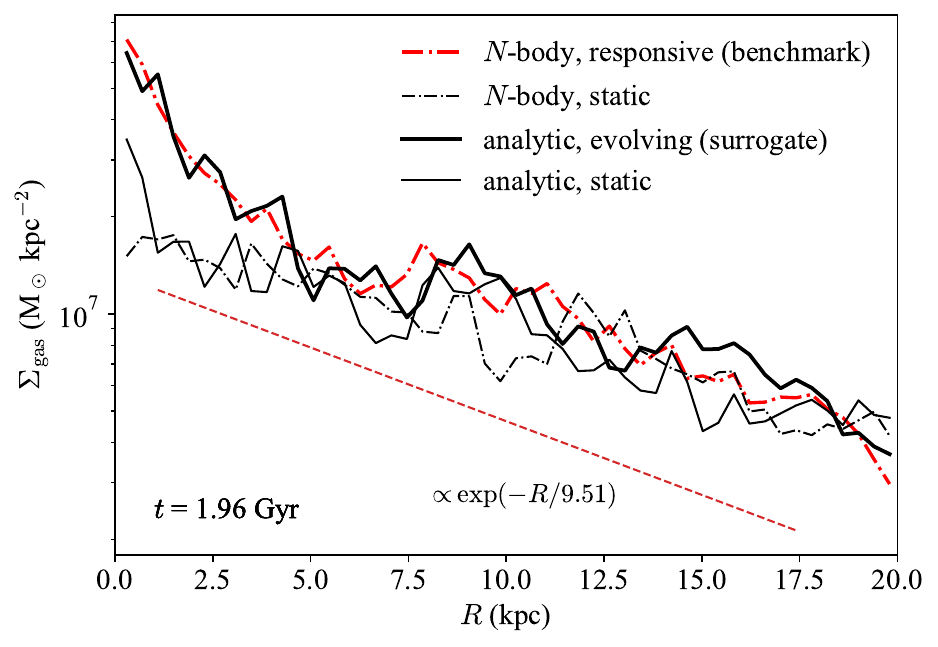 }
\includegraphics[width=\columnwidth]{ 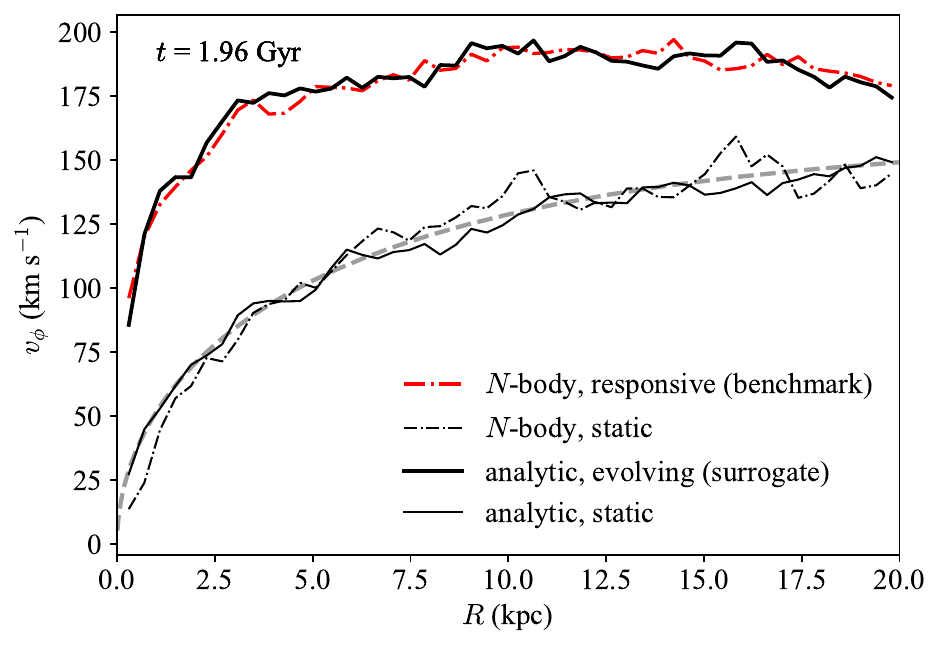 }
\caption{Structural and kinematic properties of the gas disc after $\sim 2$~Gyr of evolution. Top: Surface density. The red, dashed line corresponds to an exponential profile with scalelength $R_{\rm d} \approx 9.1$~kpc, fitted to the thick, red, dot-dashed curve in the range $R \in [ 1, 17.5 ]$~kpc, and it is included for reference only.
It corresponds to the azimuthal average as a function of cylindrical radius $R$.
Bottom: Rotational velocity around the galaxy, azimuthally averaged as a function of cylindrical radius $R$. The rotation velocity induced by the {\em initial} DM potential is shown by the thick, grey, dashed curve.}
\label{fig:disc_gas_structkin}
\end{figure}

The analysis of the properties of the stellar disc have revealed already that the disc formation can be broadly recovered in the surrogate halo run. Now we turn to the analysis of the properties of the gas disc. For brevity, given the similarly good agreement with the results presented above, we limit the presentation of the corresponding results to the two most relevant observables: the disc's density profile and its rotational velocity,  These are displayed in Fig.~\ref{fig:disc_gas_structkin}. Some relevant measurements, including the disc gas mass and the radial scalelength, are summarised in Tab.~\ref{tab:runs}.

Focusing on the top panel, it is apparent that, compared to the stellar component, the gaseous disc component is significantly more extended, with a scalelength nearly a factor $\sim3$ larger. This is true for all the models. Similarly to the results pertaining to the structure and extension of the stellar disc, the density profile of the gas disc in the surrogate halo runs broadly matches that of the benchmark run, while the profile of the gas disc that develops in either of the static halo runs displays a significant difference with respect to the latter.

Not entirely surprising, the same is true for the rotation velocity of the gas around the galaxy, displayed in the bottom panel of Fig.~\ref{fig:disc_gas_structkin}. The mean rotation profile of the gas in the benchmark run is matched by that of the gas in the surrogate halo runs, while the gas disc in the static halo runs appears to rotate on average at a significantly lower speed around the galaxy. The reason is the same as in the case of the stellar disc: The deepening of the potential well in the case of an evolving halo, which is not captured, by definition, by a static halo.

\subsection{Internal structure of the disc} \label{sec:psd}

\begin{figure}
\centering
\includegraphics[width=\columnwidth]{ 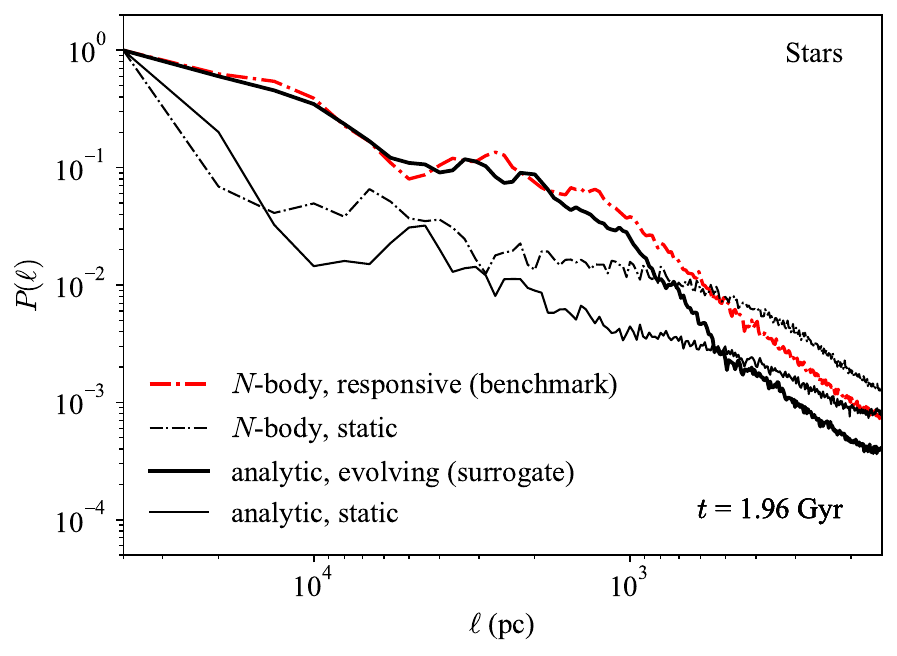 }
\includegraphics[width=\columnwidth]{ 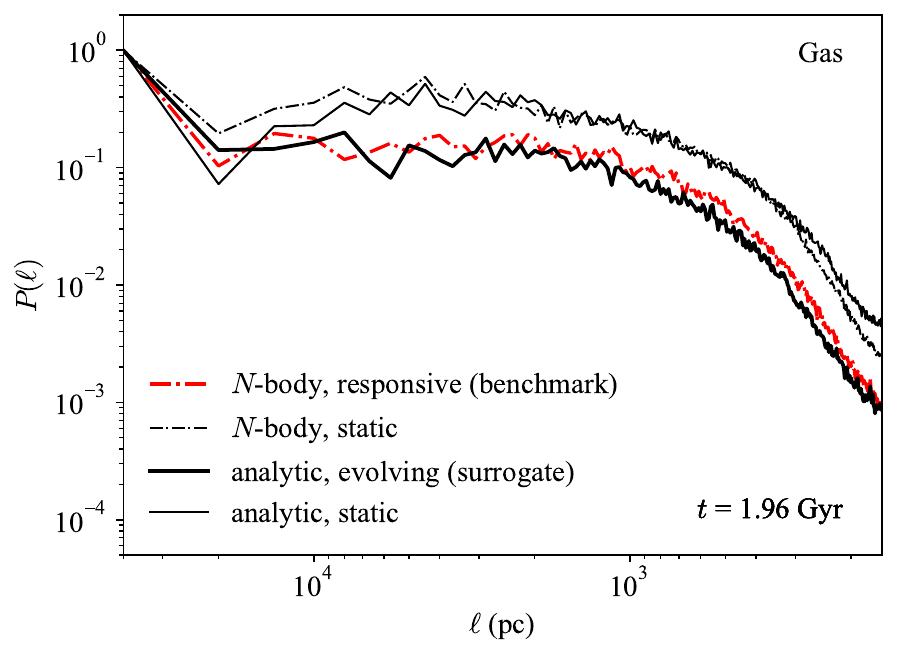 }
\caption{Power spectral density of the mass distribution in the disc after $\sim 2$~Gyr of evolution. Top: Stellar component. Bottom: Gaseous component.}
\label{fig:psd}
\end{figure}

For the purpose of a quantitative analysis of the {\em internal} disc structure, we choose the use of power spectra of the projected stellar and gas densities for each of our models. The use of this mathematical tool is a widely adopted approach to characterise and quantify the structural properties of the mass distribution within galaxies  \citep[e.g.][]{sta99a,gri17b,ren21a}. We have applied this technique for this very same purpose with success before \citep{tep25a}. The analysis of the gas surface density is particularly relevant as it provides a useful probe of the structure of the interstellar medium \citep[ISM;][]{bou10b}.

To this end, individual face-on $xy$-projections of the three-dimensional (3D) stellar and gas distributions centred on the disc over an area of \mbox{$40~\kpc\times40~\kpc$} were calculated  at the last time step (\mbox{$\sim 2$~Gyr}) by integrating the stellar and gas densities along the $z$-axis over a 40~kpc path in length, when the disc has formed and appears settled (cf. App.~\ref{app:disc_form}). Each of the stellar and gas projections, or density maps, was sampled with \mbox{$N_{\rm pix} = 501$} pixels per side, implying a pixel size of \mbox{$\sim80$~pc}, corresponding to roughly twice the limiting, spatial resolution of the AMR grid. Examples of these projections at $t \approx 2$~Gyr are displayed in the left column of Figs.~\ref{fig:disc_gas} and \ref{fig:disc_stars} for the gas and stellar components, respectively. A visual comparison already reveals that the discs have a different structure across runs (rows).

Next, the power spectrum was computed by performing a discrete Fast Fourier Transform (FFT) over each of the `raw' (i.e.~without windowing, mirroring, or zero-padding)\footnote{We have checked that mirroring, windowing, or zero-padding \citep[cf.][]{gri17b} does not noticeably affect the shape of RPS; any of these manipulations only changes its overall amplitude quasi-uniformly across all scales.
} density maps, and averaging the amplitude over the different spatial frequency ($k$) bins, weighing by the width of the frequency annulus. The radially averaged power spectrum (RPS) thus computed is a measure of the variance of the gas mass (or stellar mass) distribution at each spatial scale \mbox{$\ell = 2 \pi / k$}. To this end, we adapt the code provided by Bert Vandenbroucke.\footnote{Available at \url{https://bertvandenbroucke.netlify.app/2019/05/24/computing-a-power-spectrum-in-python/}.} 

The RPS obtained for each component and each run is displayed in Fig.~\ref{fig:psd}. The top panel shows the results for the stellar disc component, while the bottom panel shows the results for the gaseous disc component.
For each model, each RPS has been normalised to the amplitude of the initial ($t=0$) RPS at the largest scale ($l = 40$~kpc).

For any model, there are noticeable differences between the RPS of the disc's stellar component (top) and its gaseous component (bottom), which are however not unexpected given the different nature of these two components -- the gas being dissipative in contrast to the disipationless nature of the stars. Most importantly, and in line with the findings discussed above in regards to the disc's global structural and kinematic properties, the stellar and gaseous components of the disc that form within the surrogate halo each has an inner structure very similar to that of the corresponding components of the disc that forms within the fully responsive halo run. However, there are small, although noticeable differences in the small-scale power ($\ell \lesssim 10^3$~pc) of the stellar disc structure (top). We believe the small-scale power is associated with inhomogeneities in the DM distribution induced by gravitational interactions between baryons and DM near the galaxy centre, which are likely missing from the analytic reconstruction. This belief is consistent with the fact that the largest errors in the force magnitude and direction are found in the inner regions too (cf.~Sec.~\ref{sec:ver}).

In contrast, the RPS of the disc that forms within a static halo displays substantially more power in its stellar component, and less power in its gaseous component, across all scales, compared to the disc in the benchmark run, indicating a vastly different inner structure. The difference in the RPS of the gas component implies substantially different ISM structure, as mentioned above. If the benchmark run provides the more realistic representation of the ISM, this difference implies that the static halo runs produce an unrealistic ISM structure, making such setups of limited utility for galaxy formation studies.

\section{Summary} \label{sec:summ}

Our aim is to extend the use of an analytic reconstruction for the potential of a cosmologically evolving DM halo -- a surrogate halo -- to systems including a gaseous component. The ultimate goal is to simulate a standalone galaxy within a halo evolving under cosmological boundary conditions. In other words, to replay a cosmological/zoom-in simulation with a fixed halo evolution history, systematically varying the properties of the baryons that evolve within, in addition to the sub-grid physics `recipes' that govern their evolution, in addition to gravity and hydrodynamics.

As a first step towards this goal, we have setup a cooling halo configuration and simulated with it the formation of a galactic stellar disc within a fully responsive DM halo, to be used as a benchmark. We have reconstructed the evolution of the halo's potential in this simulation with an analytic function  -- a surrogate --, and we have replayed the benchmark simulation, replacing the fully responsive DM halo by its surrogate. For the purposes of this work, we have defined a validated replay as one that reproduces the key structural and kinematic properties of the disc that forms in the benchmark run -- the `benchmark disc'.

With this in mind, we have compared the total mass, the density profile (radial and vertical), the rotational velocity and velocity dispersion (radial and vertical), and the inner structure of the benchmark disc -- both its stellar and gaseous components -- with the disc that forms in the surrogate halo simulation -- the `surrogate disc'. In addition, we have run a set of control simulations, which feature either a static N-body DM halo or a static analytic halo reconstruction, and have followed the same analysis for each.

We have found that the surrogate disc matches the benchmark disc across the board of bulk-, structural- and kinematic properties of both its stellar component and its gaseous component. In stark contrast, the disc that forms the control simulations fails to reproduce the benchmark disc in every aspect.

\section{Discussion} \label{sec:concl}

Owing to the initial, net, non-zero angular momentum of the hot gas halo in the cooling halo setup, a disc forms in each of the simulations we presented, regardless of the dynamic nature of the host DM halo. This suggests that a sufficient condition for disc formation is the net, non-zero angular momentum of the gas \citep{pic11a,sal12a,kre20a}, in addition to a stable centre \citep[][]{myr26a}.

However, the structural and kinematic properties of the disc that forms in each case are substantially different. A static halo potential, that is, one fixed at $t=0$ in the simulation, leads to the formation of a disc that is too extended, too thick and, at the same time, kinematically too cold, compared to the disc that forms within an evolving DM potential. This is true regardless of whether the DM halo is represented by an (noisy) N-body model, or a smooth, analytic potential.\footnote{It must be said that it is unclear what the result would be if the baryons were evolved within a halo potential fixed at the {\em end} of the simulation, as we have not conducted this experiment.}

An analytic evolving, i.e.~surrogate, DM potential, in contrast, leads to the formation of a disc with properties astonishingly similar to those of a disc forming in the reference simulation within a fully responsive DM halo. This is striking for at least two reasons.

First, the analytic halo -- being rigid by construction -- does not respond dynamically to the baryons evolving within \citep[][for a succinct review, see \citealt{pon14a}]{blu86a,gne04a}.
In plain terms, there is no dynamical feedback loop between DM and baryons, but rather the former dictates how the latter are to evolve within the rigid dark-matter scaffold. Thus, it seems that an acceptable simulation replay is predicated on two requirements: 1) that reconstruction is that of a halo {\em that has evolved hosting the baryons}; and 2) that the reconstruction is of sufficient accuracy. And perhaps a third one: that the evolution of the halo's potential is sampled with enough cadence. The first requirement ensures that the halo's response is encoded within the analytic solution. The second and third, that this information is faithfully preserved.

Second, a rigid halo does not allow for angular momentum transfer, which in fact aids the formation of a disc, because baryons, by design, retain their initial angular momentum \citep{fal80a}, and renders the disc more stable against bar formation \citep[e.g.][]{zav08a}. The agreement between the two discs in our simulations strongly suggests that the angular momentum exchange within baryons and a fully responsive halo does not significantly affect the disc formation and its evolution, {\em as long as the disc does not feature a bar} (see Sec.~\ref{sec:cav}).

One last point to note is the relationship between disc formation and the gravitational potential's structure. It has been argued that a necessary condition for the formation of a disc is that the potential steepens as the baryons fall in \citep[][see also \citealt{sem26a} and references therein]{hop23a}. 
Yet we have shown that although the potential deepens with time, its central profile becomes progressively flatter.
The argument has been called into question by more recent work \citep{myr26a}, and our simple setup supports this challenge (see also \textcolor{blue}{Nyhagen et al., in prep.})

\subsection{Caveats} \label{sec:cav}

Given the complex nature of the mechanisms, and their intricate interplay, leading to the disc formation, the agreement between the results of the benchmark run and the surrogate halo run is truly remarkable. But the approach is not without caveats.

First, we need to consider the effect of the spatial resolution. The reader may recall that the AMR grid structure responds to two refinement criteria (cf.~Sec.~\ref{sec:bench}). One of these is based on the maximum allowed number of collisionless particles (DM, stars) in a cell. Since in the runs featuring an analytic halo model there are initially no particles (stars), it is not clear how to adjust $m_{\rm sph}$ to ensure a consistent refinement configuration across all simulations from the outset, and over the system's full evolution.
There are a few options available to us, such as adding a fixed, geometric refinement criterion (e.g.~a spherical volume with a prescribed maximum refinement level), or limiting the level of refinement triggered by particles in a cell; we shall explore these in future work. For the time being, we have found that the grid structure in the disc region in the benchmark- and surrogate halo runs is roughly comparable, and both very different from the static halo runs -- as anticipated. This is in part the reason for the stark contrast between the former and the latter.

Interestingly, a surrogate halo run identical to the one discussed earlier, but ran at a slightly lower resolution (\mbox{$m_{\rm sph} = 10^{-1}$~M$_{\rm cu}$} rather than $5\times10^{-2}$~M$_{\rm cu}$), recovers every disc observable remarkably well except for the disc mass, which ends up being roughly 20 percent lower (not shown). The latter circumstance is neither alarming, not entirely unexpected, since a lower spatial resolution naturally shifts the gas density distribution function towards the lower end, thus reducing the amount of gas above the threshold allowed to form stars. This experiment suggests that the results of the surrogate halo run are not heavily affected by resolution.

There is a subtle point to be made, though. It is not clear at the moment whether a fully responsive halo is contributing artificially to grid refinement through shot noise. If so, then adopting a strategy that leads to a reduction in refinement is not necessarily wrong, or inconsistent. In fact, the surrogate halo setup may actually be cleaner, smoother, and less noisy. The key question is whether, as in the experiment described above, now the AMR grid is under-resolving gas instabilities, disc structure, central dynamics, star formation regions, etc. We shall address these questions in future work.

The second caveat worth noting is the idealised nature of the cooling halo setup. By construction, the host DM halo -- and by extension the hot halo -- lacks any structure (triaxiality, substructure, global shape distortions), or misalignment between their angular momenta, all typical of DM halos found in cosmological N-body~/~hydrodynamical simulations \citep[e.g.][]{jin02a,bul17a}.

However, these circumstances do not, in themselves, constitute a fundamental limitation. Indeed, it is possible with our framework to setup a DM+gas configuration with some degree of angular momentum misalignment, informed by cosmological simulations. Crucially, earlier work has shown that it is possible to account for the complex structure of DM halos, their kinematics, and their environment while analytically reconstructing the effect of its potential on a system evolving within \citep[e.g.][]{low11a,san20b,aro22a}. But, as mentioned before, none of these has yet tested this (more complex) type of reconstruction on the evolution of an inner, {\em gas-dynamical} system. Thus, for the time being it is unclear how these elements would affect a similar experiment to ours; in particular, we ignore at present whether the absence of these elements may have facilitated the correspondence between the disc formation in the benchmark run and the surrogate halo run.

One potential caveat of the surrogate halo approach is the inherent absence of dynamical friction between the dark matter and the baryons, which governs angular momentum transfer between the DM halo and the disc. The latter has been shown to affect the properties (size/strength, pattern speed) of a central stellar bar \citep[][]{deb00a,ath03l,ath05i}, and to drive gas inflows and central mass growth \citep[q.v.][]{kor04c}. It is unclear to what extent bar growth and slowdown would be realistically reproduced in an externally provided halo potential, nor how nuclear star formation and central bulge development would be affected. We shall investigate these effects in future work.

Finally, it is worth noting that the surrogate halo simulation is as computationally expensive as the benchmark simulation, at least in the case of our particular setup. The surrogate halo run with {\tt rho\_ana} is in fact significantly more costly, which is not entirely unexpected, as mentioned earlier. However, we suspect that replaying a 'zoom-in' simulation would be less costly than the original simulation. This is, amongst others, the subject of future studies.

\subsection{Outlook} \label{sec:out}

The early findings reported here are encouraging,
suggesting that it is feasible to replace a fully responsive N-body DM halo by an analytic reconstruction in more complex simulations (cosmological, zoom-in), and replay the simulation many times, thus opening the door to performing a series of controlled, standalone galaxy simulations with cosmological gravity boundary conditions, each with a different set of baryonic initial conditions, sub-grid prescriptions and their parameters, etc., in order to explore their effect on the properties of the synthetic galaxy.

Some of the many experiments this approach enables include: an in-depth exploration of the conditions for disc settling (both numerical parameters and physical such as selecting a different gas angular momentum profile); testing, and potentially improving upon, the \citet{mo98a} formalism; exploring the physics for disc reformation during major mergers, just to mention a few.

We are confident that, although replaying a zoom-in simulation presents additional challenges, the problem is by no means intractable.
In the next paper of the series, we will put our formalism to the test using a MW-analogue extracted from the zoom-in, cosmological  simulation suite {\sc vintergatan} \citep{age21l,ren21b,ren21c,seg22a,rey2023}. This simulation is ideally suited, as it employs the same simulation code (\ramses) and the same sub-grid physics models \citep[star formation, stellar feedback, enrichment, cooling; see][and references therein]{age20a}, thereby minimising potential biases arising from differences in numerical methods or physical prescriptions.


\section*{Acknowledgements}

We are indebted to Eugene Vasiliev for comments on an early version of this manuscript and for his continued support with \agama, and to Romain Teyssier for his continued help with \ramses.
This work has been supported by the Australian Research Council (ARC) through grant ``Galactic seismology: a new window on Milky Way's evolution'' (ID: DP220103384; CI: Bland-Hawthorn). OA acknowledges support from the Knut and Alice Wallenberg Foundation, the Swedish Research Council (grant 2025-04892), the Swedish National Space Agency (SNSA grants 2023-00164 and 2025-00405), the LMK foundation, and eSSENCE, a Swedish strategic research programme in e-Science. CF acknowledges funding provided by the Australian Research Council (Discovery Projects DP230102280 and DP250101526), and the Australia-Germany Joint Research Cooperation Scheme (UA-DAAD).
We further acknowledge high-performance computing resources provided by the Australian National Computational Infrastructure (NCI; grants af49, ek9) and the Pawsey Supercomputing Centre (project~pawsey0810) awarded through the NCI-Sydney Scheme, the National Computational Merit Allocation Scheme, and the ANU Merit Allocation Scheme, and resources provided by the Leibniz Rechenzentrum and by the Gauss Centre for Supercomputing (grants~pr32lo, pr48pi and GCS Large-scale project~10391).

TTG acknowledges the use of OpenAI's ChatGPT (GPT-5.5) to assist in the generation and refinement of code {\em snippets} used in this analysis, as well as with language editing and improving the clarity of selected portions of the manuscript. All generated code was reviewed, validated, and, where necessary, modified by the authors prior to use. The scientific content, analysis, and interpretations remain those of the authors.


All figures and movie frames were created with Matplotlib \citep{hun07a}, and all animations assembled with {\small FFmpeg}.\footnote{\url{http://www.ffmpeg.org} }
This research has made use of NASA's Astrophysics Data System (ADS) Bibliographic Services.\footnote{\url{http://adsabs.harvard.edu} }

\section*{DATA AVAILABILITY}

The software and data underlying this article will be shared on reasonable request to the corresponding author.


\appendix

\section{Disc assembly and dynamical settling} \label{app:disc_form}

\begin{figure}
\centering
\includegraphics[ width=0.75\columnwidth ]{ 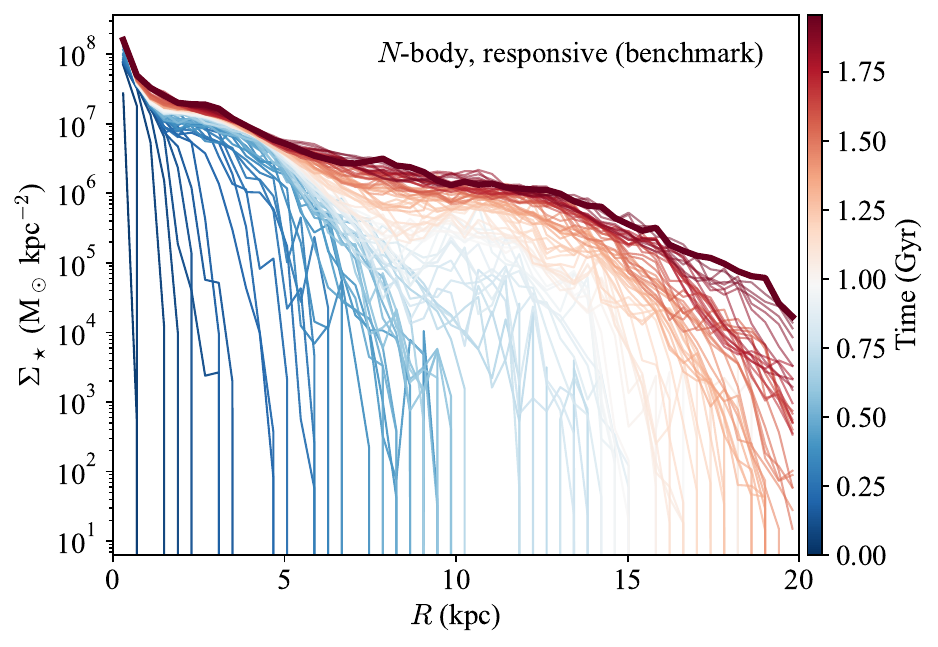 }
\includegraphics[ width=0.75\columnwidth ]{ 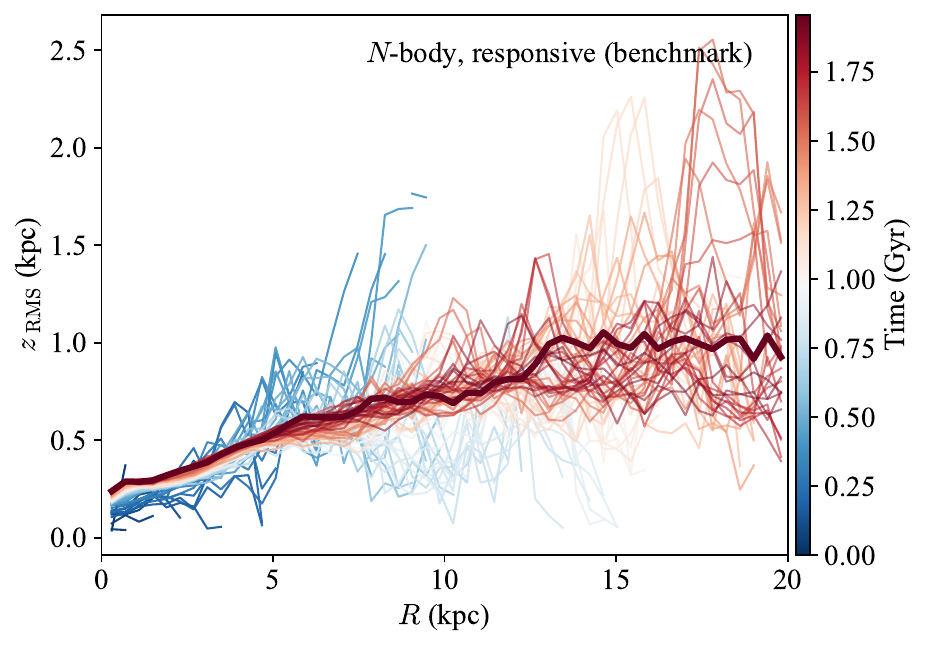 }
\includegraphics[ width=0.75\columnwidth ]{ 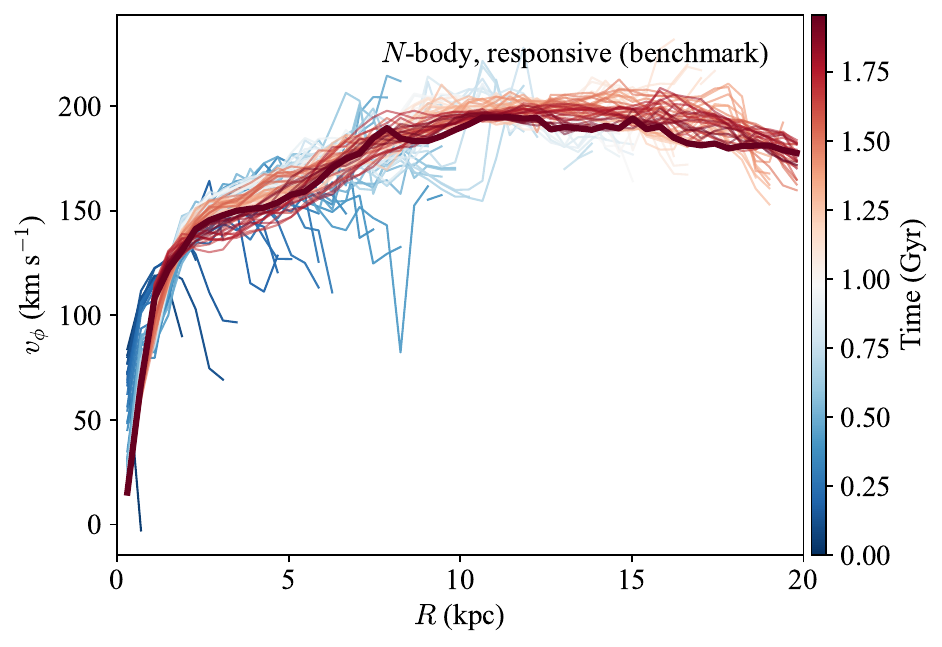 }
\includegraphics[ width=0.75\columnwidth ]{ 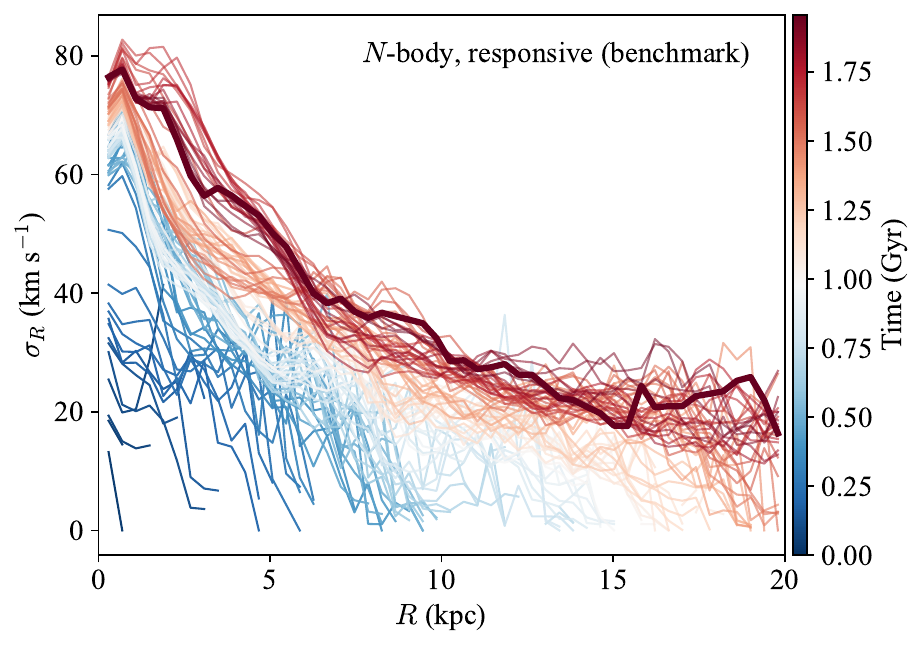 }
\includegraphics[ width=0.75\columnwidth ]{ 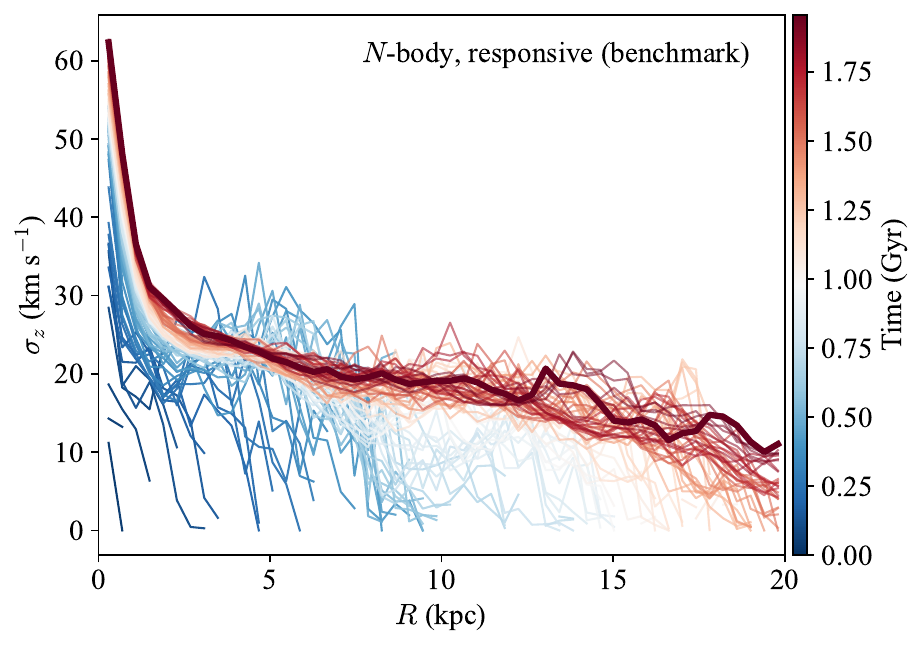 }
\caption[  ]{ Evolution of the stellar disc's properties (benchmark run). From top to bottom: Stellar surface density; RMS height above the mid-plane; rotational velocity around the centre; radial- and vertical velocity dispersion. The final state of the disc in each case is highlighted by a thick, dark line-style.}
\label{fig:dform}
\end{figure}

Fig.~\ref{fig:dform} displays the evolution of several structural and kinematical properties of the stellar disc in the benchmark run. Each of these observables corresponds to an azimuthally averaged, radial profile. From top to bottom, the quantities shown are: stellar surface density; root-mean square height above the mid-plane; rotational velocity around the galaxy; radial velocity dispersion; and vertical velocity dispersion. Each panel displays a series of curves, one for each time step $\delta t \approx 10$~Myr, over an evolution timespan of $\sim 2$~Gyr. The final state of the disc, i.e.~at $t \approx 2$~Gyr, in each case is highlighted by a thick, dark line-style. Note that the latter are identical to the red, thick, dot-dashed line-style displayed in Figs.~\ref{fig:disc_star_struct} - \ref{fig:disc_star_kin}. The apparently slower change towards the end of the timespan, compared to the dramatic changes apparent at the beginning of the evolution, suggests that the disc is rapidly converging towards a stationary state. Indeed, the disc appears well-formed and dynamically settled by the end of the 2~Gyr period. This is important in regards to the comparison of the disc evolution in the benchmark run with the other runs over (and by the end of) the same timespan.

\section{Higher multipole orders} \label{app:lmax}

\begin{figure}
\centering
\includegraphics[width=\columnwidth]{ 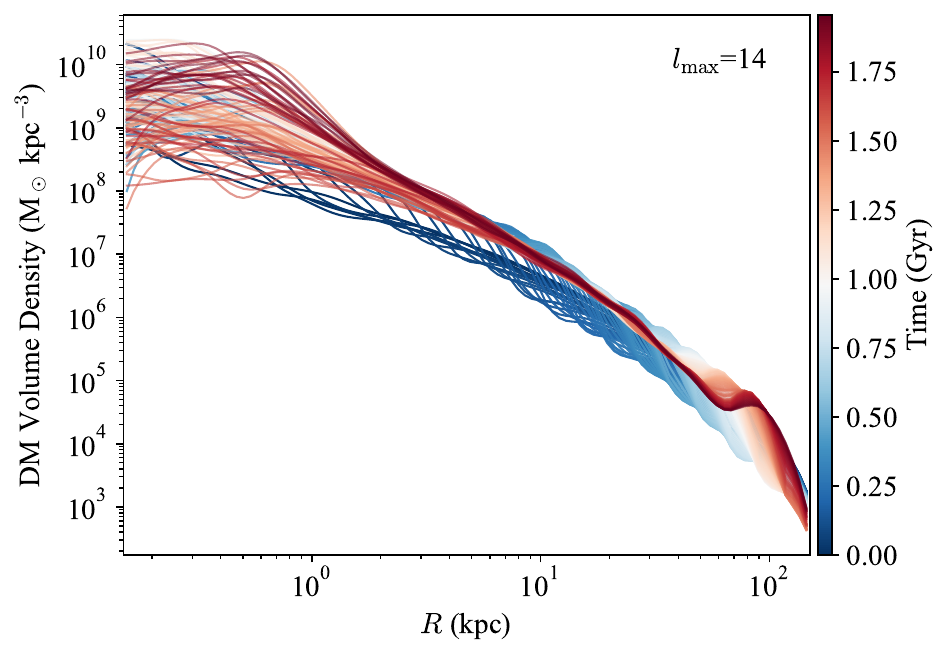 }
\includegraphics[width=\columnwidth]{ 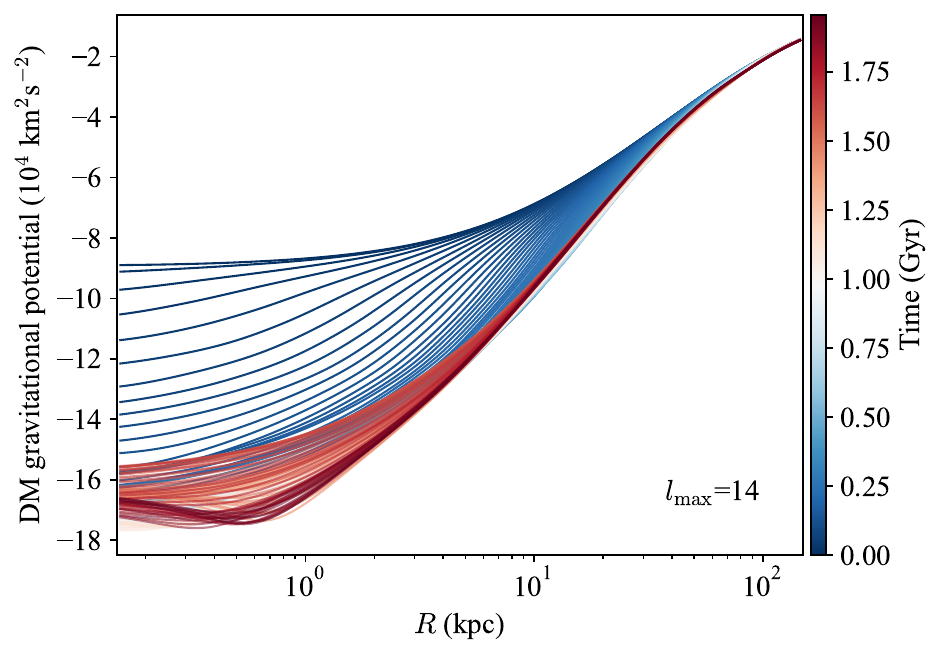 }
\caption[  ]{ Reconstructed density (top) and reconstructed potential (bottom) of the DM component in the N-body responsive halo model, adopting \mbox{$\ell_{\rm max} = 14$}. The profiles for {$\ell_{\rm max} = 20$} are comparable, and therefore, omitted. }
\label{fig:phi_ana2}
\end{figure}

In addition to the fiducial value of the maximum multipole order, $\ell_{\rm max} = 6$, we have used $\ell_{\rm max} = 14$ and $\ell_{\rm max} = 20$ in the reconstruction of the DM halo potential and its evolution in the benchmark run. Fig.~\ref{fig:phi_ana2} displays the density profile (top) and the potential profile (bottom) for $\ell_{\rm max} = 14$. While not identical to the reconstruction using $\ell_{\rm max} = 6$ (cf. Fig.~\ref{fig:phi_ana}), they are similar. The same is true for the reconstruction using $\ell_{\rm max} = 20$, which, to avoid redundancy, is not shown here. Given this similarity, we expect that a surrogate halo simulation using the reconstruction with $\ell_{\rm max} = 14$ (or $\ell_{\rm max} = 20$) to deliver similar results to our fiducial run with $\ell_{\rm max} = 6$.

\section{Smaller opening angles} \label{app:theta}

\begin{figure*}
\centering
\includegraphics[width=0.32\textwidth]{ 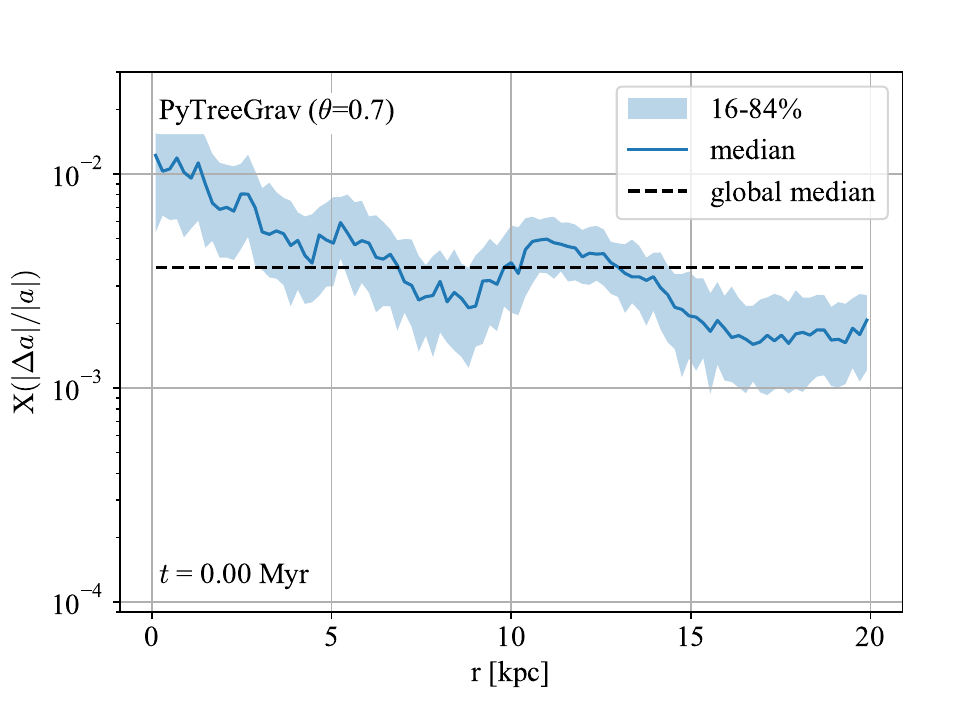 }
\includegraphics[width=0.32\textwidth]{ 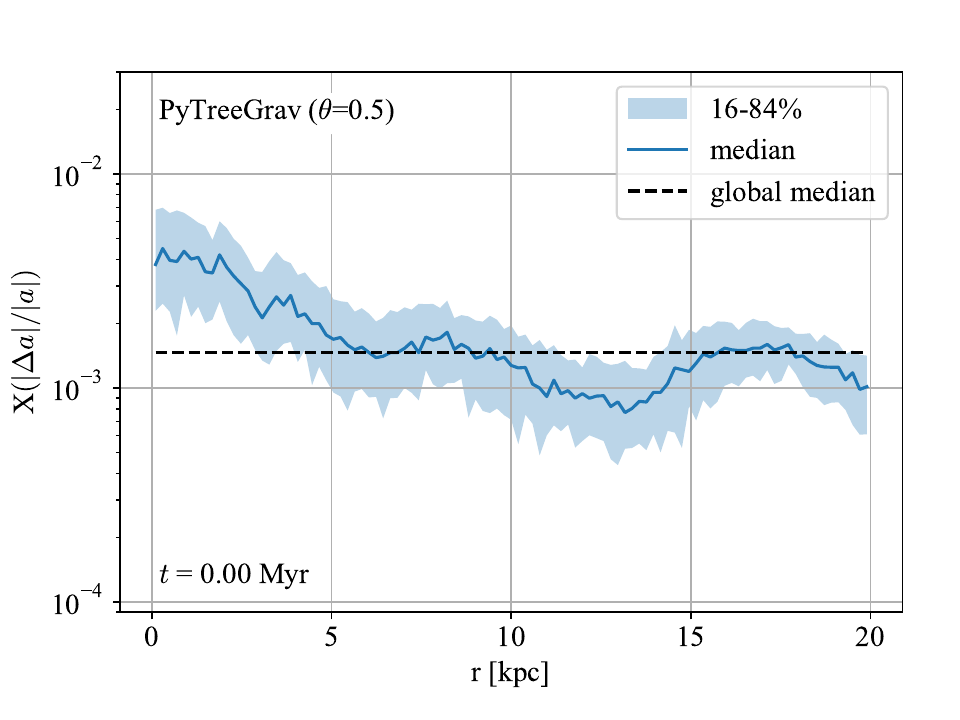 }
\includegraphics[width=0.32\textwidth]{ 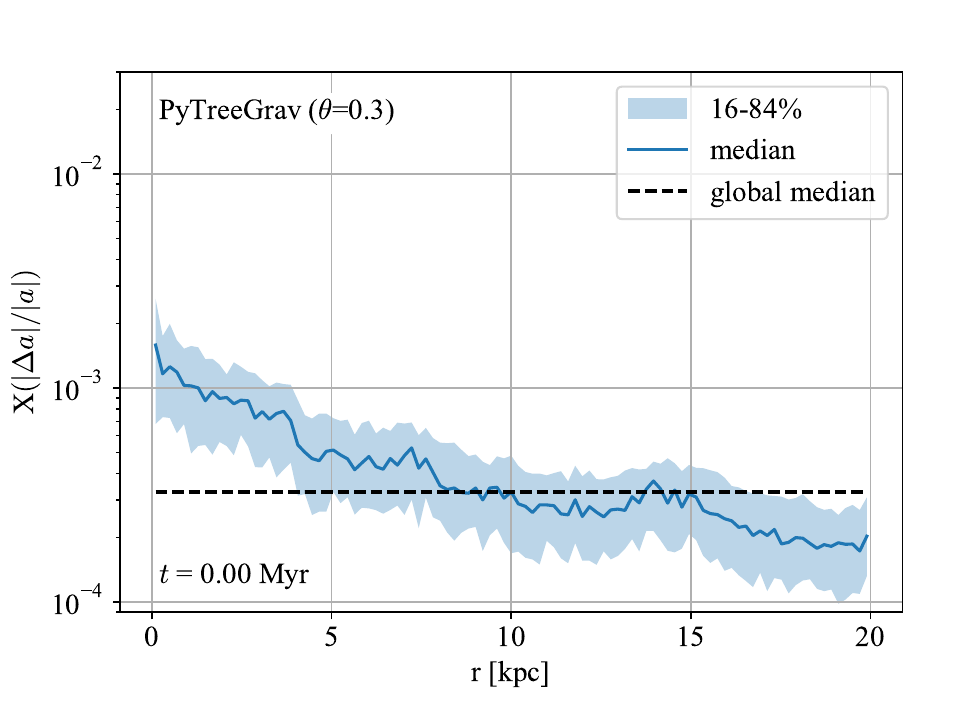 }\\
\includegraphics[width=0.32\textwidth]{ 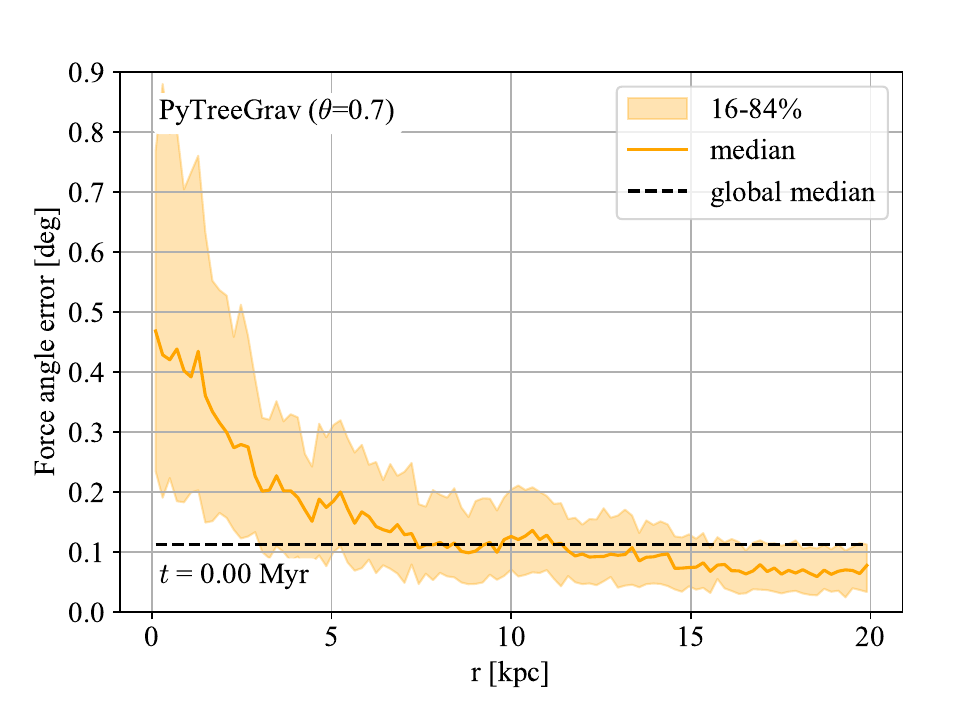 }
\includegraphics[width=0.32\textwidth]{ 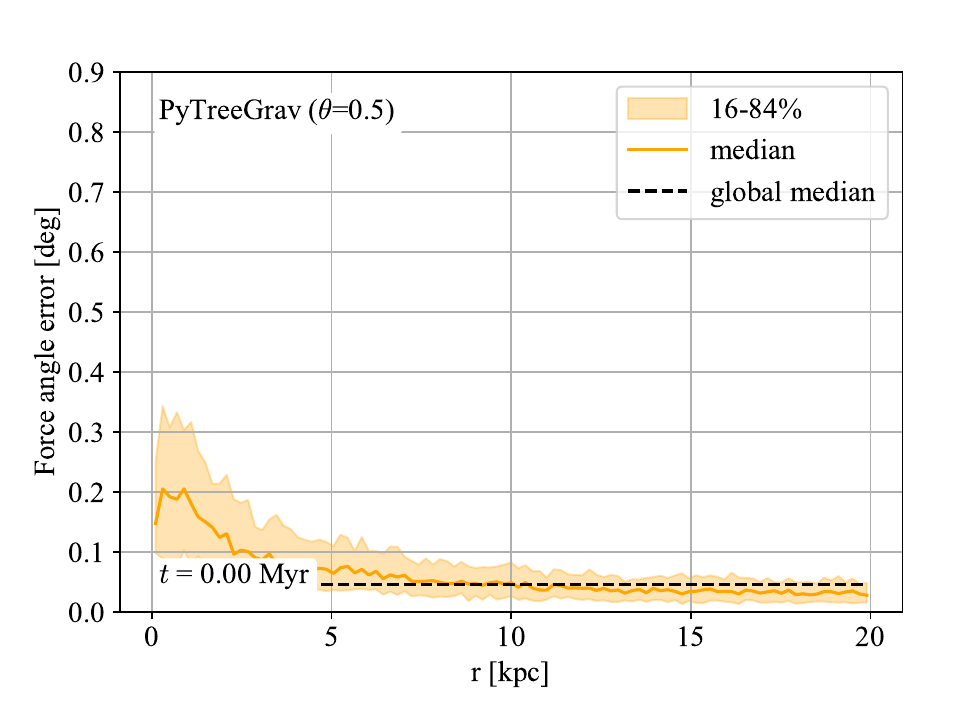 }
\includegraphics[width=0.32\textwidth]{ 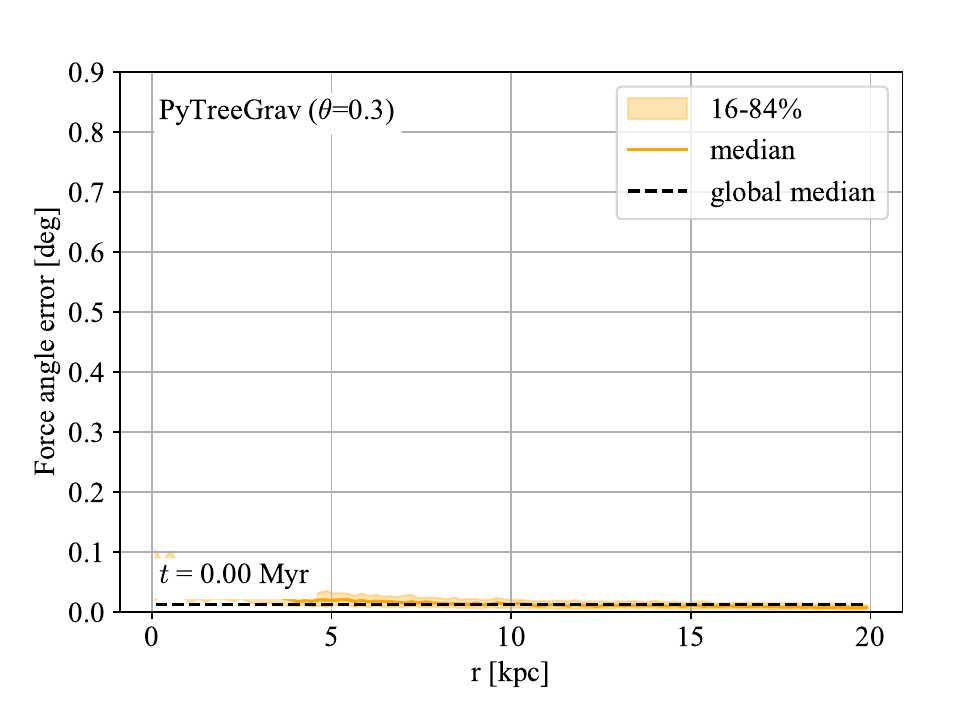 }
\caption[  ]{ Potential reconstruction with pytreegrav \citep{gru21a}. Top: Relative force magnitude error. Bottom: Force misalignment, Columns from left to right correspond to $\theta = 0.7$, 0.5, and 0.3, respectively. Note that the scale along the $y$-axis in each row is intentionally preserved across columns.}
\label{fig:rec}
\end{figure*}

We test the accuracy of the potential reconstruction with {\sc pytreegrav} \citep{gru21a}. To this end, we construct a potential by exact direct summation (`brute force') of $N$ dark matter particles. The specific potential induced at position $\vec{r}$ is (note that the particles have all the same mass $m$),
$$
	\Phi(\vec{r}) = - G m \sum_{i=1}^N \frac{1}{ | \vec{r} - \vec{r_i} | } \,
$$
and regard this as the `true' potential. We then reconstruct the potential with a tree method adopting different values of the opening angle $\theta \in \{0.3, 0.5, 0.7 \}$; recall that lower $\theta$ values improve the accuracy of the reconstruction, but they may increase the computational cost. We test each of these potential estimates against the true potential using the relative force magnitude error and the force misalignment as metrics (see Sec.~\ref{sec:rec}). The results are displayed in Fig.~\ref{fig:rec}. The upshot is that, as expected, the lower the value of $\theta$, the lower the magnitude error and the better the alignment; a reconstruction using $\theta = 0.3$ yields near perfect agreement with the true potential, with an average median magnitude error of order $5\times10^{-4}$, and an average misalignment close to 0, across all probed radii. But even $\theta = 0.7$ yields very satisfactory agreement with the true N-body potential, with an average median magnitude error of order $5\times10^{-3}$, and an average misalignment of less than 0.5 deg, thus justifying our  assumption adopted in the analysis that this particular reconstruction be regarded as the `true' potential.\\

\begin{figure}
\centering
\includegraphics[width=\columnwidth]{ 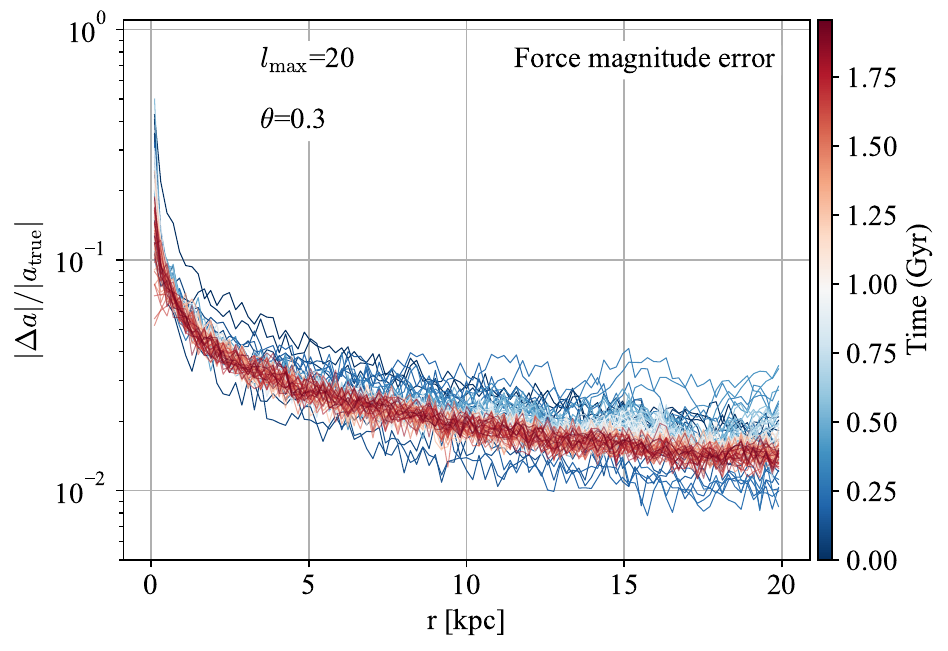 }
\includegraphics[width=\columnwidth]{ 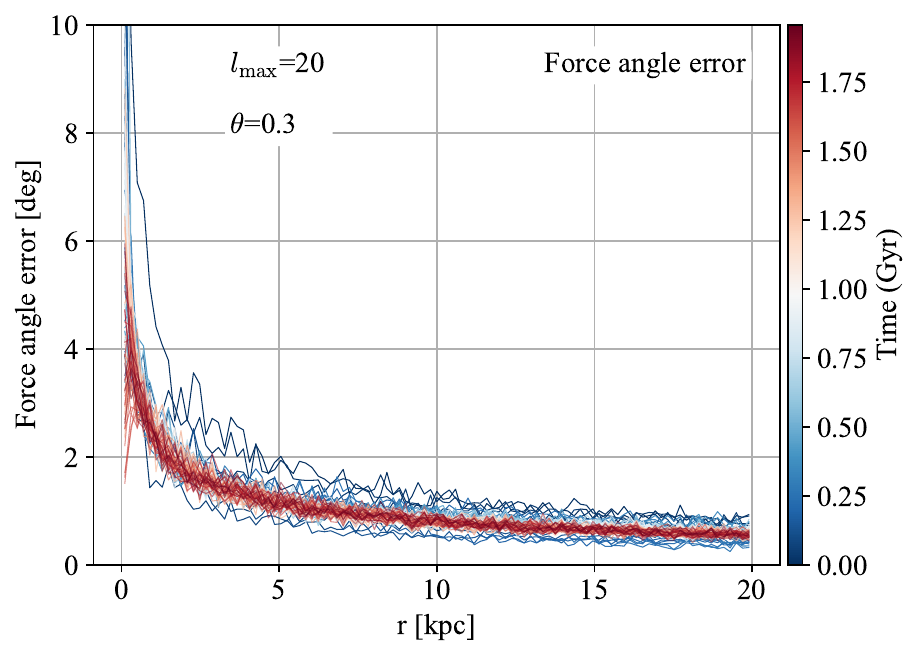 }
\caption[  ]{ Evaluating the potential reconstruction. Top: Relative force magnitude error. Middle: Force directional error.
Note that an angle error of $\lesssim 10$~deg implies that the force vectors are aligned to within 2 percent.  }
\label{fig:eval2}
\end{figure}

The effect of increasing the multipole order to $\ell_{\rm max} = 20$ and decreasing the opening angle to $\theta = 0.3$, compared to our fiducial values, on the force reconstruction is showcased in Fig.~\ref{fig:eval2}. Compared to the results displayed in Fig.~\ref{fig:eval}, we find no appreciable improvement. This suggests that the potential reconstruction using our fiducial values $\ell_{\rm max} = 6$ and $\theta = 0.7$ appear to be of sufficient quality.

\bibliographystyle{mnras} 
\bibliography{complete,joss,oscar} 

\bsp	
\label{lastpage}
\end{document}